\documentclass[a4paper,10pt]{article}
\usepackage[utf8]{inputenc}
\usepackage{amsmath}
\usepackage{amssymb} 
\usepackage{hyperref}
\usepackage{color,soul}
\usepackage{lscape}
\usepackage{graphicx}
\usepackage[square,sort&compress,comma,numbers]{natbib}

\def\lsim{\raise0.3ex\hbox{$\;<$\kern-0.75em\raise-1.1ex\hbox{$\sim\;$}}}
\def\gsim{\raise0.3ex\hbox{$\;>$\kern-0.75em\raise-1.1ex\hbox{$\sim\;$}}}

\definecolor{Lila}{rgb}{0.5,0.,1}

\newcommand{\gev}{\mbox{ GeV}}

\newcommand{\mev}{\mbox{ MeV}}

\newcommand{\fb}{\mbox{ fb}}
\newcommand{\BR}{{\rm BR}}
\newcommand{\tgg}{\to \gamma\gamma}
\newcommand{\taa}{\to A_1 A_1}

\graphicspath{{figs/}}
\title{{The NMSSM lives: with the $750$~GeV diphoton excess}}
\author{Florian Domingo$^{a,b}$, Sven Heinemeyer$^{a,b}$, Jong Soo Kim$^{a}$, Krzysztof Rolbiecki$^{a,c}$}
\date{$^{a}${\em Instituto de F\'isica Te\'orica (UAM/CSIC), Universidad Aut\'onoma de Madrid, Cantoblanco, E-28049 Madrid, Spain}\\[.4em]
$^{b}${\em Instituto de F\'isica de Cantabria (CSIC-UC), E-39005 Santander, Spain}\\[.4em]
$^{c}${\em Institute of Theoretical Physics, University of Warsaw, PL-02093 Warsaw, Poland }}

\begin{document}

\maketitle
\vspace{-6.5cm}\rightline{IFT-UAM/CSIC-16-019}
\rightline{arXiv:1602.07691 [hep-ph]}\vspace{6.5cm}
\begin{abstract}
We propose an NMSSM scenario that can explain the excess in the diphoton
spectrum at $750$~GeV recently observed by ATLAS and CMS. We show that
in a certain limit with a very light pseudoscalar one can reproduce the
experimental results without invoking exotic matter. The $750$~GeV
excess is produced by two resonant heavy Higgs bosons with masses
${\sim}750$~GeV, which subsequently decay to two light
pseudoscalars. Each of these decays to collimated photon pairs that
appear as a single photon in the electromagnetic calorimeter. A mass gap
between heavy Higgses mimics a large width of the $750$~GeV peak. The
production mechanism, containing a strong component via initial $b$
quarks, ameliorates a possible
tension with $8$~TeV data compared to other production modes. We also
discuss other constraints, in particular from low-energy
experiments. Finally, we discuss possible methods that could distinguish
our proposal from other physics models describing the diphoton excess
in the Run-II of the LHC.
\end{abstract}


\section{Introduction}
The ATLAS~\cite{ATLASdiphoton2015} and CMS~\cite{CMSdiphoton2015} experiments at the Large Hadron Collider (LHC) have both reported an excess in the diphoton channel at an 
invariant mass of about $750$~GeV corresponding to a local (global) significance of $3.6\,\sigma$ ($2.0\,\sigma$) and $2.6\,\sigma$ ($1.2\,\sigma$), respectively. 
The result is of course not conclusive, but if the excess were confirmed, this would be the first sign of new physics at the terascale energies.

The simplest explanation requires the production of an $s$-channel spin-0 or spin-2 resonance according to the Yang--Landau theorem~\cite{Landau:1948kw,Yang:1950rg}. The 
observed cross section of roughly $\mathcal{O}(10)$~fb is relatively large and thus it is natural to assume that the new resonance is produced via the strong interaction and 
have a large decay rate into diphotons. A light quark initiated resonance would be in severe tension with the LHC Run-I, since the parton luminosity ratio between 
$\sqrt{s}=13$~TeV and $8$~TeV is relatively small for light quark initial states. As a consequence, the resonance would have been observed at LHC Run-I if it were produced via 
quark--antiquark initial states. For a gluon induced resonance the tension with $8$~TeV is reduced but still significant~\cite{Kim:2015ksf}. On the contrary, associated 
production with $b$ quarks does not suffer from $8$~TeV constraints. Moreover, the reported event topology is consistent with the single production of a resonance, i.e.\ 
non-resonant production of the $750$~GeV particle in a cascade decay
\cite{Kim:2015ron} is disfavored since no additional activity was observed in
the peak-region events.%
\footnote{However, the heavy parent resonance scenario can still be
  phenomenologically viable if the lighter resonance mainly decays into
  dark matter~\cite{DeRomeri:2016xpb}.}%
~Finally, the apparently 
large width of around 45~GeV, preferred by ATLAS, points to large couplings to its daughter particles. However, strict constraints exist on decays of heavy resonances into 
electroweak gauge bosons and light Standard Model (SM) fermions and thus the resonance should decay into final states which evade all current experimental searches, implying e.g.\ a 
large invisible decay rate or decays to quarks and gluons. Another way out (which we actually consider in this paper) would be the presence of two overlapping resonances with 
narrow widths which allow to explain the large width within the current experimental accuracy~\cite{Cao:2016cok}. 

The observed diphoton rate cannot be explained with a SM-like Higgs boson because its tree-level decays into third generation quarks and/or gauge bosons are too large compared 
to the loop induced decays into diphoton final states. However, simple extensions of the SM Higgs sector such as a singlet extension or Two-Higgs-Doublet Model (2HDM) are also 
plagued with too small diphoton rates and the way out is to introduce new vector-like fermions: see e.g.\ \cite{Angelescu:2015uiz,Falkowski:2015swt}; see \cite{Staub:2016dxq} for an overview.
There are only a few phenomenologically viable explanations within the framework of supersymmetry (SUSY)~\cite{Drees:2004jm}. It seems to be impossible to find a solution within 
the Minimal Supersymmetric Standard Model
(MSSM)~\cite{Gupta:2015zzs}. New vector-like fermions and singlets have
to be added to the particle spectrum of the MSSM in order to explain the
diphoton excess~\cite{King:2016wep,Tang:2015eko,Hall:2015xds,Ma:2015xmf,Wang:2015omi}. Other
SUSY solutions either involves $R$-parity
violation~\cite{Allanach:2015ixl,Ding:2015rxx} or assume a very low SUSY breaking
scale~\cite{Bellazzini:2015nxw,Petersson:2015mkr,Casas:2015blx}.

The Next-to-Minimal Supersymmetric Standard Model (NMSSM) -- see~\cite{NMSSM} for a review -- is a well-motivated supersymmetry-inspired extension of the 
SM. Beyond the elegant features of supersymmetric models in view of the hierarchy problem or one-step unification, and their potential in 
terms of Dark Matter (DM), the original purpose of the NMSSM rests with the `$\mu$-problem'~\cite{Kim:1983dt} of the simpler MSSM: this issue is addressed 
via the addition of a singlet superfield to the matter content of the MSSM, the `$\mu$'-parameter then being generated dynamically when the singlet takes a 
vacuum expectation value (v.e.v.). Additionally, the NMSSM has received renewed attention ever since the Higgs discovery in the Run-I phase of the LHC~\cite{HiggsDisco}, 
due to its interesting features in terms of a supersymmetric interpretation of the observed Higgs signals -- see~\cite{Domingo:2015eea}
for a recent analysis and list of references. While several versions of the NMSSM can be formulated, we will focus here on the simplest one, characterized by a $Z_3$-symmetry 
and CP-conservation. Let us stress here that we will not include new exotic matter but rely strictly on the simple matter content of this model.

\begin{figure}[t]
\begin{center}
\includegraphics[clip, trim=2.5cm 17cm 0.cm 2.5cm,width=0.7\textwidth]{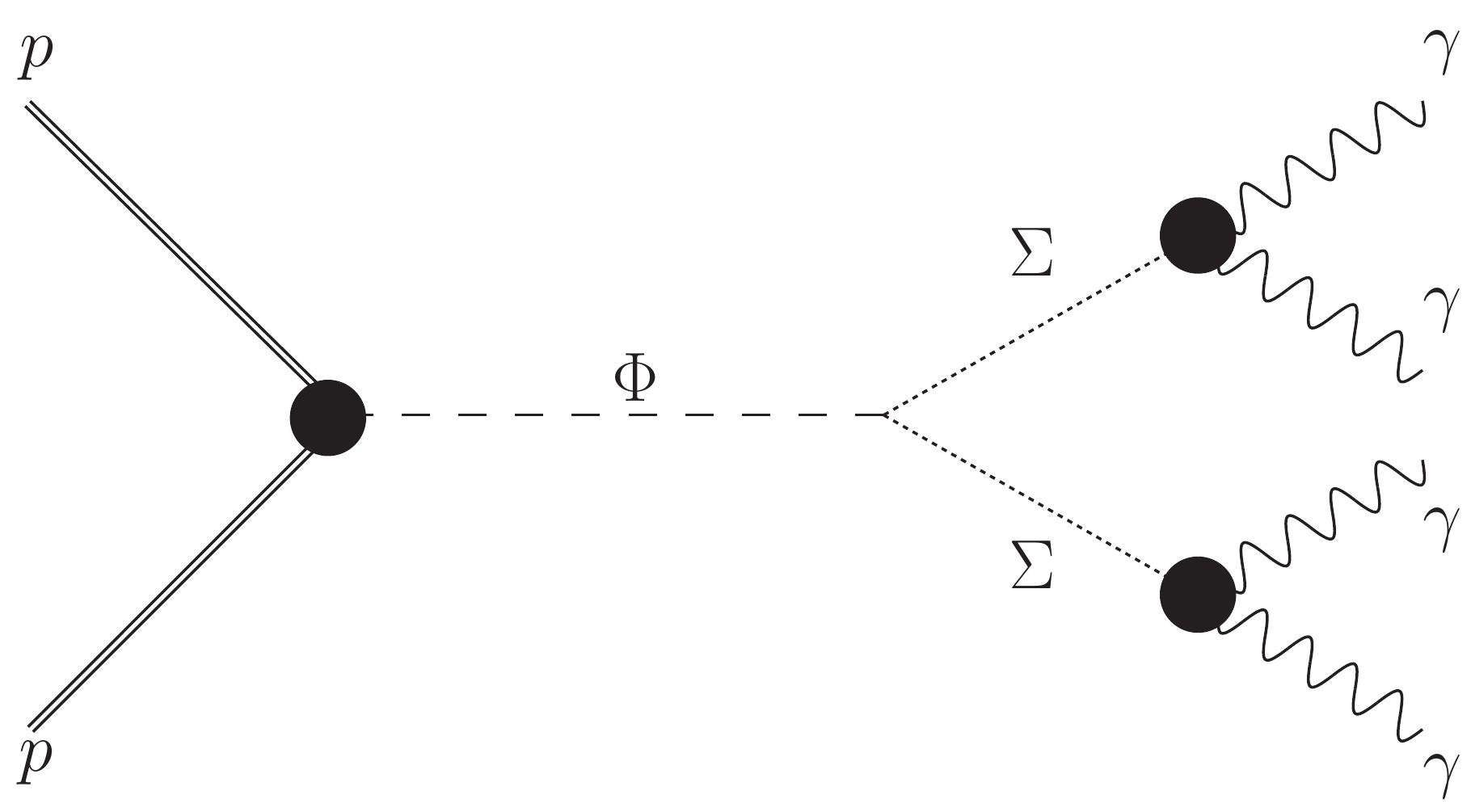}
\caption{The resonant production of $\Phi$ followed by the decay to two $\Sigma$ scalars and photons. The final-state photons are pairwise 
collimated. 
\label{fig:process}}
\end{center}
\end{figure}

Our purpose in this paper is to present a phenomenologically viable scenario accounting for the diphoton excess at ${\sim}750$~GeV in the context of the NMSSM. This explanation 
rests on the possibility that a $pp\to \Phi\to2(\Sigma\to\gamma\gamma)$ process -- see Figure~\ref{fig:process} -- could not be distinguished from a diphoton signal in the 
experimental searches~\cite{Bi:2015lcf,Knapen:2015dap,Agrawal:2015dbf,Dobrescu:2000jt,Larios:2002ha,Chang:2006bw,Chang:2015sdy}. The NMSSM Higgs sector then offers a suitable framework to embed this topology: $\Sigma$ can be identified with a very light CP-odd 
singlet decaying dominantly into a diphoton pair, while heavy CP-even doublet and the CP-even singlet have to be combined to mimic a $750$~GeV resonance with adequate properties. 
Note that contrarily to the proposals which we mentioned earlier, the mechanism that we consider does not rely on the ad hoc inclusion of additional matter (e.g.\ vector-like 
fermions) and uses only the existing features and degrees of freedom of a
rather simple and well-motivated model.%
\footnote{We observe also that this mechanism can be transposed 
to an apparently simpler -- in fact theoretically less {motivated} --
singlet extension of the (Type II) 2HDM.} 
While our project was in its finalizing stages, we became 
aware of another recent proposal to explain the diphoton signals within
the NMSSM~\cite{Ellwanger:2016qax}, which shares some traits with our
interpretation but also differs in several respects. The diphoton
decay of the pseudoscalar in~\cite{Ellwanger:2016qax} relies on a
substantial `quasi-mixing' of this Higgs state with the $\eta$-meson:
this  
requirement induces substantial limits from $\Upsilon$-decays and results in a quite constrained regime. In our case, we consider the mixing with the $\pi^0$ and estimate this 
effect more quantitatively using the chiral perturbation theory for pions. Moreover, we shall propose a wider selection of benchmark points, illustrating the flexibility of the 
mechanisms that we employ. Furthermore, we shall analyze in further detail how our scenario compares to the experimental data and study complementary signatures.

In the following section, we will detail how the NMSSM provides an acceptable framework for the ${\sim}750$~GeV signal and address several phenomenological issues which constrain 
the parameter space. We will thus propose specific examples of NMSSM points satisfying these requirements before discussing their relevance in fitting the diphoton excess 
in Section~\ref{sec:collider}. Then we consider other experimental signatures
that are specific to our model and have promising prospects in the Run-II,
including the 
decays of the light pseudoscalar to $e^+e^-$ pairs. We also discuss
possible experimental signatures from the higgsino, slepton, and heavy Higgs bosons sector. We will conclude with a brief
summary in the last section.


\section{Embedding the \boldmath{$750$}~GeV diphoton excess
  in the parameter space of the NMSSM} 

The NMSSM Higgs sector -- see e.g.\ \cite{NMSSM} -- consists of two doublets, $H_u=(H_u^+,H_u^0)^T$ and $H_d=(H_d^0,H_d^-)^T$, coupling in a Type II fashion 
at tree level, as well as a singlet $S$. Once the Goldstone bosons are rotated away, one is left with a pair of charged states $H^{\pm}$, two doublet and one 
singlet CP-even degrees of freedom, $h_u$, $h_d$, and $h_s$, and finally one doublet and one singlet CP-odd components, $A_D$ and $A_S$. The simplest, $Z_3$- and 
CP-conserving version of the NMSSM, which we are considering in this paper, counts seven parameters in the Higgs sector, which can be denoted as $m^2_{H_u}$, 
$m^2_{H_d}$, $m^2_{S}$ -- three soft squared masses, $\lambda$, $\kappa$ -- two Yukawa-like supersymmetric couplings, $A_{\lambda}$ and $A_{\kappa}$ -- the two 
corresponding trilinear soft couplings. It is customary to use the minimization conditions of the Higgs potential to trade three of these parameters, e.g.\ the soft 
squared masses, for the Higgs v.e.v.'s, $v_u=\left<H_u^0\right>=v\sin\beta$, $v_d=\left<H_d^0\right>=v\cos\beta$ (so that $\tan\beta\equiv v_u/v_d$) and 
$s=\left<S\right>$. The electroweak properties then imply the identification $v=\left(2\sqrt{2}G_F\right)^{-1/2}\simeq174$~GeV, $G_F$ representing the Fermi constant. 
Moreover, in order to provide a more physical handle on the parameter space, we define $\mu\equiv\lambda s$ -- which 
is analogous to its MSSM `$\mu$' counterpart and also sets the tree-level mass of the higgsino states -- and $M_A^2\equiv\frac{2\lambda s}{\sin{2\beta}}\left(A_{\lambda}+\kappa s\right)$  -- which sets the scale 
of an approximate Higgs $SU(2)$-doublet, generally heavy, i.e.\ $M_A> 125$~GeV, if one wants to avoid the complications of light non-SM-like doublet 
states.\footnote{A similar quantity would be the mass of the charged-Higgs state: $m^2_{H^{\pm}}=M_A^2+M_W^2-\lambda^2v^2$ at tree level. We will prefer $M_A$ in the following
since it leads to slightly simpler expressions.} In the following, we 
shall employ the parameter set $(M_A,\tan\beta,\mu,\lambda,\kappa,A_{\kappa})$.


\subsection{Identifying the light state `\boldmath{$\Sigma$}'}

In order to fit the diphoton excess, we want to employ the $pp\to \Phi\to2(\Sigma\to\gamma\gamma)$ topology. Obviously, $\Phi$ would be a CP-even (in order to allow a decay 
to two lighter identical states) Higgs state at ${\sim}750$~GeV, while $\Sigma$ could be in principle CP-odd or CP-even: given the limited 
pool of Higgs states in the NMSSM and the fact that we will have another use for the CP-even singlet (see below), this state will be identified, however, with 
the singlet-like pseudoscalar. For its diphoton decay to be
indistinguishable from a single photon, this Higgs state would have
to be light enough: $m_{\Sigma}\lsim0.5 \gev$~\cite{Agrawal:2015dbf}. 
We observe that this scenario with a light CP-odd state is phenomenologically viable, in view of the current status of collider constraints. However, 
severe limits intervene at low mass from, e.g.\ flavor observables or the properties of the SM-like Higgs boson $H_{\rm SM}$ -- a sizable branching fraction 
$H_{\rm SM}\to\Sigma\Sigma$ would be at odds with the measurements of ATLAS and CMS in the LHC Run-I phase. Note that the $\Sigma$ state cannot be dominantly doublet in 
view of the consequences for the spectrum: this would indeed imply the presence of a light charged Higgs (likely in contradiction with limits from top decays \cite{topH}) and a light 
CP-even doublet state (in tension with LEP \cite{Barate:2003sz} and likely to open up sizable unconventional decays of $H_{\rm SM}$).

We complete this discussion with a remark concerning the naturalness of a light CP-odd Higgs in the NMSSM: in two specific limits, this particle appears as the pseudo-Goldstone 
boson of an approximate symmetry of the Higgs sector:
\begin{itemize}
 \item for $\kappa\to0$, i.e.\ $\kappa\ll\lambda$, the Higgs potential is invariant under a $U(1)$ Peccei--Quinn symmetry, under which the charges 
$Q^{\rm PQ}_{H_u,H_d,S}$ of the doublets and singlet satisfy $Q^{\rm PQ}_S+Q^{\rm PQ}_{H_u}+Q^{\rm PQ}_{H_d}=0$;
 \item for $A_{\kappa},A_{\lambda}\to0$, one obtains another $U(1)$ symmetry, with the Higgs charges $Q^{R}_{H_u,H_d,S}$ satisfying $2Q^{R}_S-Q^{R}_{H_u}-Q^{R}_{H_d}=0$:
this is the $R$-symmetry limit since the Higgs charges for the specific choice $Q^{R}_S=-2/3$ coincide with those that these fields would receive under a 
genuine $R$-symmetry (also broken by the gaugino masses) of the NMSSM with unbroken supersymmetry. Note that with our choice of parameters, 
$A_{\lambda}=-\frac{\kappa}{\lambda}\mu\left(1-\frac{\lambda M_A^2}{2\kappa\mu^2}\sin{2\beta}\right)$ (by definition of $M_A$).
\end{itemize}
We will see that, in the scenario under consideration, the factor $\left(1-\frac{\lambda M_A^2}{2\kappa\mu^2}\sin{2\beta}\right)$ has to be small, so that the $R$-symmetry 
limit can be invoked

Most of the characteristics of the NMSSM which intervene in the interpretation of the signal at ${\sim}750$~GeV and the subsequent constraints can be understood
from the relations at tree level. This follows from the fact that the relevant physics is driven by comparatively heavy or singlet-like states, for which the 
radiative corrections are relatively mild in proportion. We shall therefore propose a discussion at tree level in the following. Note, however that loop 
corrections play a crucial role for the mass of the SM-like Higgs state so that we will employ tools including leading radiative effects in the numerical 
analysis (see below). Of course, the exact correlations at tree level are slightly displaced by loop corrections so that small adjustments in the choice of parameters
will prove necessary and the relations that we derive below should not be understood as rigid constraints, but rather as qualitative guidelines/trends.

In the base $(A_D,A_S)$, the tree-level squared-mass matrix for the NMSSM CP-odd sector reads
\begin{multline}
 {\cal M}^2_P=\begin{bmatrix} M_A^2 & -3\kappa\mu v\left(1-\frac{\lambda}{6\kappa}\frac{M_A^2}{\mu^2}\sin{2\beta}\right)\\
 -3\kappa\mu v\left(1-\frac{\lambda}{6\kappa}\frac{M_A^2}{\mu^2}\sin{2\beta}\right) & m^2_{A_S}
 \end{bmatrix}\hspace{0.5cm},\\ m^2_{A_S}\equiv3\frac{\kappa}{\lambda}\mu\left[-A_{\kappa}+\frac{\lambda^2v^2}{2\mu}\sin{2\beta}\left(1+\frac{\lambda}{6\kappa}\frac{M_A^2}{\mu^2}
\sin{2\beta}\right)\right]\,.
\end{multline}
We observe that the singlet squared-mass $m^2_{A_S}$ is largely determined by the choice of $A_{\kappa}$ (since the other terms are small). In particular, low values of 
$A_{\kappa}$ ensure that the singlet is light. Note also that in this case $M_A^2\gg m^2_{A_S}$ -- as we will see later, $M_A\sim750$~GeV. The subdominant doublet component of 
the light CP-odd mass state $A_1=\sqrt{1-P_d^2}\,A_S+P_d\,A_D\simeq A_S+P_d\,A_D$ can be obtained as approximately:
\begin{equation}\label{Pd}
 P_d\simeq\frac{3\kappa\mu v}{M_A^2}\left(1-\frac{\lambda}{6\kappa}\frac{M_A^2}{\mu^2}\sin{2\beta}\right)\,.
\end{equation}
From now on, we identify this light state $A_1$ to the state $\Sigma$ of the $pp\to \Phi\to2(\Sigma\to\gamma\gamma)$ topology.

Now we will address the decays of $A_1$. For the mass range under consideration, $m_{A_1}\lsim0.5$~GeV, the interplay of the pseudoscalar
with the strongly interacting sector is non-trivial. In a naive partonic approach, the diphoton decay is significantly suppressed -- $\mathcal{O}(10^{-5})$ -- for a usual 
NMSSM pseudoscalar, as long as the competition of the hadronic and muonic final states remains open. Then, below three times the pion mass (the decay to two pions would violate parity), hadronic final states seem inaccessible 
so that the pseudoscalar would then mainly decay to muons or, below the $\mu^+\mu^-$ threshold, to $\gamma\gamma$ -- $\mathcal{O}(70\%)$ at ${\sim}210$~MeV, and the $e^+e^-$ 
final state would eventually dominate at lower masses ($\lsim160$~MeV). Moreover, in such a regime, the pseudoscalar would appear as relatively long lived below the 
dimuon threshold: the diphoton and $e^+e^-$ final states are essentially mediated by the small doublet component of $A_1$, resulting in the following (approximate) 
widths (for $m_{A_1}\lsim210$~MeV):
\begin{align}
 &\Gamma[A_ 1\to e^+e^-]\simeq(1.6\cdot10^{-13})\,m_{A_1}\,(P_d\tan\beta)^2\;,\\
 &\Gamma[A_ 1\to \gamma\gamma]\simeq(3\cdot10^{-12}~\mbox{GeV}^{-2})\,m_{A_1}^3\left[1+\left(\frac{m_{A_1}}{0.195\,{\small\rm GeV}}\right)^6\right]\,(P_d\tan\beta)^2.\nonumber
\end{align}
The corresponding total width is thus of order $\Gamma_{A_1}\sim(10^{-13}~\mbox{GeV})\,P_d^2\tan^2\beta\sim10^{-14}~\mbox{GeV}$ for $P_d\sim3\cdot10^{-2}$, $\tan\beta\sim10$ and 
$m_{A_1}\sim200$~MeV, and narrower at lower masses. The decay length of $A_1$ at rest is then of order $2~\mbox{cm}$. Including the boost factor in the considered 
topology would lead to a decay length of about $40$~m at the LHC, i.e.\ to an invisible pseudoscalar.

This picture is over-simplistic, however. As \cite{Ellwanger:2016qax} already noticed, hadronic effects could substantially affect the decays of $A_1$. If one were to disregard the 
tripion threshold for strong-interacting decays, the $s\bar{s}$ and $gg$ final states would completely dominate the partonic decays of $A_1$, which highlights the sensitivity of
the pseudoscalar to the strong sector. In particular, considering the couplings of the pseudoscalar to mesons, one can write the following operators:
\begin{equation}
 -{\cal L}_{\rm eff}\ni\delta m^2_{A_1\pi^0}\,A_1\pi^0+\delta m^2_{A_1\eta}\,A_1\eta+\ldots
\end{equation}
Such terms induce a mixing of the Higgs pseudoscalars with the pseudoscalar mesons or, in other words, open the possibility for $A_1\to(\pi^0)^*$ or $A_1\to(\eta)^*$ decays
(the latter being the choice of \cite{Ellwanger:2016qax}).
The magnitude of the mixing elements $\delta m^2_{A_1\pi^0/\eta}$ can be assessed by rewriting the partonic couplings of $A_1$ in terms of the axial-flavor currents and their
expression in the pion model (chiral perturbation theory)~\cite{GellMann:1968rz,Chang:2008np,Masjuan:2015cjl}:
\begin{align}
 {\cal L}&\ni-\frac{\imath\,P_d}{\sqrt{2}v}\left\{m_u\tan^{-1}\beta\,\bar{u}\gamma_5u+m_d\tan\beta\,\bar{d}\gamma_5d+m_s\tan\beta\,\bar{s}\gamma_5s\right\}A_1\nonumber\\
 &=-\frac{P_d}{2\sqrt{2}v}\partial_{\mu}\left\{\tan^{-1}\beta\,\bar{u}\gamma^{\mu}\gamma_5u+\tan\beta\,\bar{d}\gamma^{\mu}\gamma_5d+\tan\beta\,\bar{s}\gamma^{\mu}\gamma_5s\right\}A_1\\
 &=-\frac{P_d}{4v}\left\{\sqrt{\frac{2}{3}}(\tan^{-1}\beta+2\tan\beta)\,\partial_{\mu}J_{A\,1}^{\mu}+(\tan^{-1}\beta-\tan\beta)\,\partial_{\mu}J_{A\,3}^{\mu}+\frac{1}{\sqrt{3}}(\tan^{-1}\beta-\tan\beta)\partial_{\mu}J_{A\,8}^{\mu}\right\}A_1\nonumber\\
 &=-\frac{P_d}{4v}\left\{\sqrt{\frac{2}{3}}(\tan^{-1}\beta+2\tan\beta)f_{\eta_1}m^2_{\eta_1}\,\eta_1+(\tan^{-1}\beta-\tan\beta)f_{\pi}\left[m^2_{\pi}\,\pi_3+\frac{m^2_{\eta}}{\sqrt{3}}\,\pi_8\right]\right\}A_1\nonumber\;,
\end{align}
where $J_{A\,1}^{\mu}=\frac{1}{\sqrt{3}}(\bar{u}\gamma^{\mu}\gamma_5u+\bar{d}\gamma^{\mu}\gamma_5d+\bar{s}\gamma^{\mu}\gamma_5s)$, 
$J_{A\,3}^{\mu}=\frac{1}{\sqrt{2}}(\bar{u}\gamma^{\mu}\gamma_5u-\bar{d}\gamma^{\mu}\gamma_5d)$, 
$J_{A\,8}^{\mu}=\frac{1}{\sqrt{6}}(\bar{u}\gamma^{\mu}\gamma_5u+\bar{d}\gamma^{\mu}\gamma_5d-2\bar{s}\gamma^{\mu}\gamma_5s)$ and the divergences of these currents in the pion
model are determined by the equations of motion: $\partial_{\mu}J_{A\,1}^{\mu}=f_{\eta_1}m^2_{\eta_1}\,\eta_1$, $\partial_{\mu}J_{A\,3}^{\mu}=f_{\pi}m^2_{\pi}\,\pi_3$ and 
$\partial_{\mu}J_{A\,8}^{\mu}=f_{\pi}m^2_{\eta}\,\pi_8$. Here, $\eta_1$ denotes the Goldstone field associated with the $U(1)$ axial-flavor symmetry while $\pi_3$ and $\pi_8$
correspond to the generators $\lambda_3\equiv\frac{1}{\sqrt{2}}\mbox{diag}[1,-1,0]$ and $\lambda_8\equiv\frac{1}{\sqrt{6}}\mbox{diag}[1,1,-2]$ of the $SU(3)$ axial-flavor
symmetry. Dismissing the refinements of the $\pi_3$, $\pi_8$, and $\eta_1$ mixings, we can identify $\pi_3\sim\pi^0$, $\pi_8\sim\eta$ and $\eta_1\sim\eta'$. This determines:
\begin{equation}
 \delta m^2_{A_1\pi^0}=\frac{f_{\pi}}{4v}P_d(\tan^{-1}\beta-\tan\beta)m^2_{\pi}\;,\hspace{1.5cm}\delta m^2_{A_1\eta}=\frac{f_{\pi}}{4\sqrt{3}v}P_d(\tan^{-1}\beta-\tan\beta)m^2_{\eta}\;.
\end{equation}

\begin{figure}[htb]
\begin{center}
\includegraphics[width=0.48\textwidth]{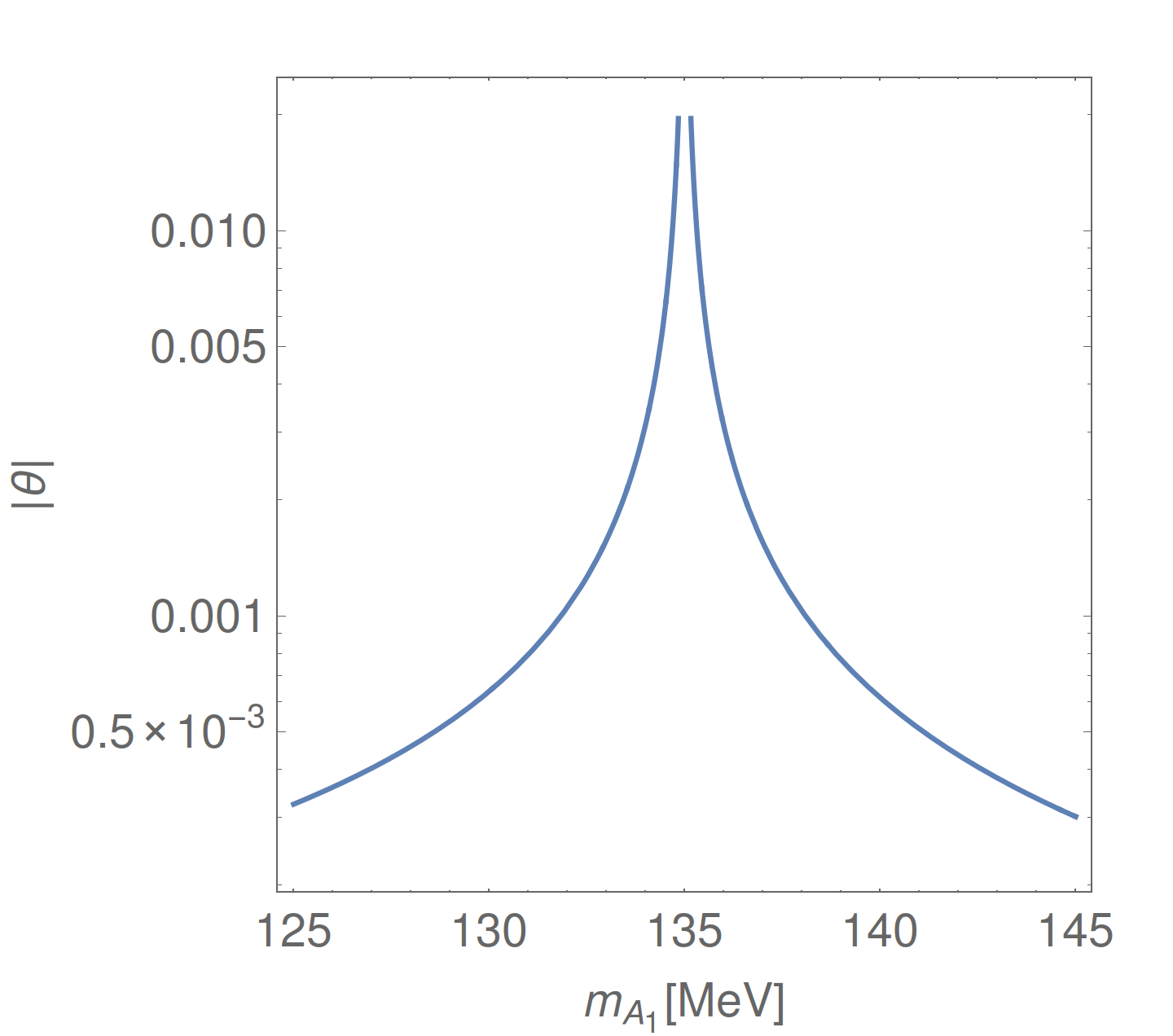}\qquad
\includegraphics[width=0.42\textwidth]{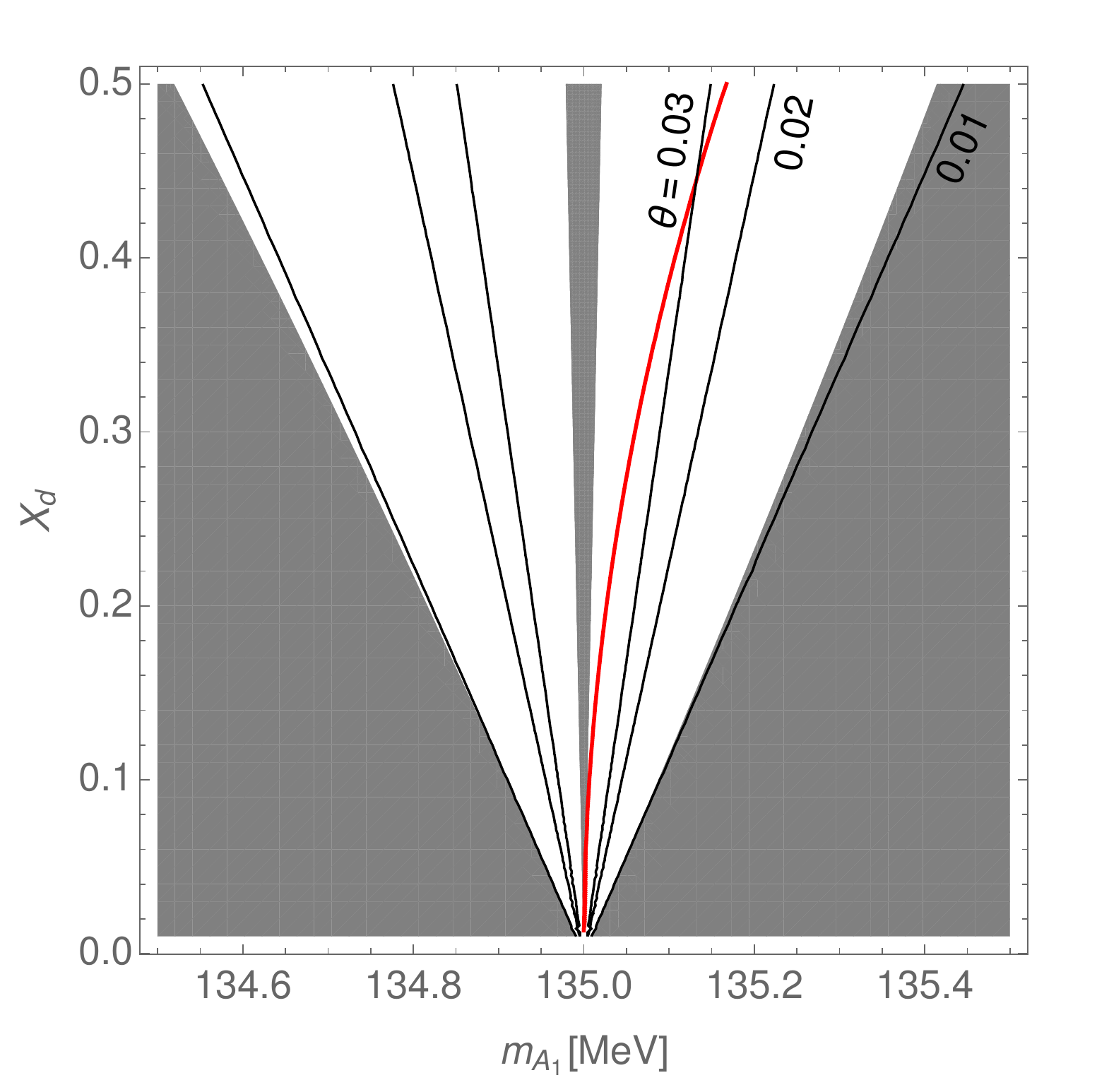}
\caption{Left: the mixing angle $|\theta|$ as a function of $A_1$ mass for
  $P_d=0.035$ and $\tan\beta=10$. Right: the contours (black) of the mixing
  angle $\theta= 0.01,\, 0.02,\, 0.03$ (from the most outer) in the
  $m_{A_1}$--$X_d$ plane  where $X_d =P_d(\tan\beta -1/\tan\beta)$ and
  $\tan\beta=10$. The gray-shaded regions 
  are excluded; see the text for details. The red contour denotes the value of
  mixing required to recover the experimental central value for $\BR[\pi^0 \to e^+e^-]$ (hence resolve the discrepancy). 
\label{fig:mixing}} 
\end{center}
\end{figure}

We observe that these mass-mixing parameters are small: with $f_{\pi}\simeq93$~MeV and the typical values $P_d\sim0.03$ and $\tan\beta\sim10$, 
$\delta m^2_{A_1\pi^0}\sim(4\cdot10^{-5}m^2_{\pi})$ and $\delta m^2_{A_1\eta}\sim(2\cdot10^{-5}m^2_{\eta})$. Therefore, they intervene only in the immediate mass-vicinity of the 
mesons, and a mixing between $A_1$ and the mesons of order
$10^{-2}$ thus requires a proximity in mass at the MeV level. In
particular, following our analysis, the decay width mediated by a  
$\eta$-state for $m_{A_1}\sim510$~MeV, as considered in \cite{Ellwanger:2016qax}, should be completely superseded by e.g.\ muon decays of $A_1$. On the other hand, multi-pion 
decays are still kinematically open for $m_{A_1}\sim510$~MeV, which may result in multi-photon final-states. We shall not elaborate further on this possibility and we now focus 
on the mixing of $A_1$ with $\pi^0$, i.e.\ assume $m_{A_1}\sim135$~MeV.
At first order in $\delta
m^2_{A_1\pi^0}$, the mixing angle $\theta$ reads $\theta\simeq\delta m^2_{A_1\pi^0}/(m_{\pi}^2-m_{A_1}^2)$ and the mass shift driven by mixing is $\frac{m_{\pi}
\theta^2}{2}\left|1-\frac{m_{A_1}^2}{m_{\pi}^2}\right|$: for a mixing of order $\theta\sim10^{-2}$
it translates into a (completely negligible) sub-keV correction. A
  sample plot of the mixing angle as a function of the $A_1$ mass for fixed $P_d$ is
  shown in the left panel of Fig.~\ref{fig:mixing}.
As an example, for $P_d = 0.035$, $\tan\beta \simeq 10$ and 
$|m_{A_1} - m_{\pi}| < 0.3 \mev$ we find $\theta = 10^{-2}$.
The consequences for the decays of $A_1$ are
sizable: via its small $\pi^0$ component, the pseudoscalar  
acquires pion-like decays with values about
\begin{equation}
\Gamma_{A_1}^{\pi}\simeq \theta^2\,\Gamma_{\pi}\sim8\cdot10^{-13}~\mbox{GeV} \;,
\end{equation}
essentially to $\gamma\gamma$ ($\BR[\pi^0\to\gamma\gamma]\sim99\%$),
which overrule the `pure-Higgs' decays of order
$10^{-15}$~GeV. The decay length in the lab frame for
the process of Fig.~\ref{fig:process}  
is then reduced to ${\sim}0.5$~m, placing most of the time
the decay vertex to photons before $A_1$ reaches electromagnetic
calorimeter.%
\footnote{We observe that for such decay lengths, the configuration of ATLAS with a ECAL detector farther from the beam could make this experiment 
more sensitive to the process than the more compact CMS, which could lead
  to a slightly weaker signal in CMS (roughly in agreement with the current
  experimental situation).} Even shorter decay lengths are possible for larger
$A_1$--$\pi^0$~mixing. For the pion, on the other hand, the
consequences are minimal as its 
natural decays are slightly decreased at the level of  
$0.01\%$ (far below the theoretical or experimental precisions) and
slightly shifted by a Higgs-like width at the level of
$10^{-16}$--$10^{-17}$~GeV (via interference terms). The
2\,$\sigma$ experimental error on the pion decay width to two photons
of $4\%$ defines our upper limit on the
mixing.\footnote{It should be noted that the measured value of $\Gamma[\pi^0\to \gamma\gamma]$ is slightly below the most recent theoretical calculations; see~\cite{Bernstein:2011bx}. 
We note that the mixing discussed here could help to explain this
tension. We speculate that the experimental setup similar to the one used for
a direct measurement of the pion decay length~\cite{Atherton:1985av}, could
also detect $A_1$ once the detectors are moved farther away from the
target. By observing a change in the measured mean decay length one could
directly probe parameters of $A_1$--$\pi^0$ mixing.} In the right panel of Fig.~\ref{fig:mixing} we show contours of the constant
mixing angle $\theta$ in the $m_{A_1}$--$X_d$ plane. With the above
constraints, the regions shaded in gray are excluded. The central part
around $m_{A_1} \simeq 135$~MeV is excluded by the upper limit of the
measurement of the width $\Gamma[\pi^0 \to\gamma\gamma]$, while the outer part
by $A_1$ decay length at the LHC, which would be larger than 
${\sim }0.5$~m.  It should be noted that the  
Particle Data Group~\cite{PDG} reports 
$\BR[\pi^0\to e^+e^-]=(6.46\pm0.33)\cdot10^{-8}$ leading to a width 
$\Gamma[\pi^0\to e^+e^-]\simeq5\cdot 10^{-16}$~GeV, known at the $6\%$ 
level. Inclusion of radiative corrections shifts this value to 
$\BR[\pi^0\to e^+e^-]=(7.48\pm0.38)\cdot10^{-8}$~\cite{Abouzaid:2006kk}, 
while the recent theoretical calculation gives 
$(6.2\pm 0.1)\cdot10^{-8}$~\cite{Dorokhov:2007bd}, so that the inclusion of 
new physics at this level may even be welcome: see
e.g.\ \cite{Kahn:2007ru}. In our case, the red contour in
Fig.~\ref{fig:mixing} denotes the mixing angles that could account for the
discrepancy in $\pi^0\to e^+e^-$ (considering the central values only).

In the following, we will assume that this mixing mechanism between $A_1$ and $\pi^0$ is responsible for the apparent $A_1$ width -- resulting in a decay length of order 
$\lsim0.5$~m for the topology of Fig.~\ref{fig:process} -- and a $\BR[A_1\tgg]\simeq99\%$.


\subsection{Identifying the heavy state `\boldmath{$\Phi$}'}

We now turn to the CP-even sector. In the base 
$(H_{\rm SM}=\cos\beta\,h_d+\sin\beta\,h_u,H_D=-\sin\beta\,h_d+\cos\beta\,h_u,H_S=h_s)$,
the tree-level squared-mass matrix reads
\begin{multline}\label{M2S}
 {\cal M}^2_S=\begin{bmatrix} M_Z^2\cos^2{2\beta}+\lambda^2v^2\sin^2{2\beta} & (\lambda^2v^2-M_Z^2)\sin{2\beta}\cos{2\beta} & 2\lambda\mu v\left[1-\left(\frac{M_A\sin{2\beta}}{2\mu}\right)^2\right]\\
(\lambda^2v^2-M_Z^2)\sin{2\beta}\cos{2\beta} & M_A^2+(M_Z^2-\lambda^2v^2)\sin^2{2\beta} & -\frac{\lambda v}{2\mu}M_A^2\sin{2\beta}\cos{2\beta}\\
2\lambda\mu v\left[1-\left(\frac{M_A\sin{2\beta}}{2\mu}\right)^2\right] & -\frac{\lambda v}{2\mu}M_A^2\sin{2\beta}\cos{2\beta} & m^2_{H_S}
 \end{bmatrix}\hspace{0.5cm},\\
m^2_{H_S}\equiv\left(\frac{2\kappa}{\lambda}\mu\right)^2\left[1+\frac{\lambda A_{\kappa}}{4\kappa\mu}\right]-\frac{\kappa\lambda}{2}v^2\sin{2\beta}\left[1-\frac{\lambda M_A^2}{\kappa\mu^2}\right]\;.
\end{multline}
It should be noted that, as long as $M_A\gg v$, the doublet sector is approximately diagonal and that the mass of the heavy doublet state $H_D$ falls close to $M_A$. Moreover, 
keeping in mind that $A_{\kappa}$ is small in order to ensure a 
light $A_1$, we observe that the mass of the CP-even singlet $H_S$ is dominated by the term $\frac{2\kappa}{\lambda}\mu$, as long as this quantity is larger than
${\cal O}(v)$. In such a regime, the lightest CP-even Higgs state $H_1$
can thus be identified with $H_{\rm SM}$ (approximately): 
$H_1\simeq\sqrt{1-S_{13}^2}\,H_{\rm SM}+ S_{13}\,H_S$, 
where the singlet component is very small: $S_{13}\ll1$.
Consequently, the $H_1$ behaves SM-like, in agreement with the
experimental results~\cite{higgs-comb}.

We now want to identify the state $\Phi$ of the $pp\to \Phi\to2(\Sigma\to\gamma\gamma)$ topology with one of the CP-even states of the NMSSM. The associated mass should 
fall close to ${\sim}750$~GeV. Both $H_D$ and $H_S$ are potential candidates. However, the $\Phi$ state should be sizably produced at the LHC, which implies
a significant coupling to SM particles, such as gluons or quarks. In the case of a dominantly singlet state, such couplings are typically suppressed in
proportion to the small doublet component of the state. On the other hand, $H_D$ has sizable couplings to tops (though $\tan^{-1}\beta$ suppressed
with respect to the SM) and to bottom quarks ($\tan\beta$-enhanced), so that a measurable production in gluon--gluon fusion ($gg$f) or in association with $b$'s ($bbh$) is
plausible. We observe that the production cross section of a $750$~GeV $H_D$ in $gg$f at $13$~TeV is well approximated by $(0.6~\mbox{pb})/\tan^2\beta$ for $\tan\beta\lsim15$,
and the $bbh$ cross section, by $(4\cdot10^{-4}~\mbox{pb})\times\tan^2\beta$. The production cross section would thus point toward an identification $\Phi\sim H_D$, which would imply 
$M_A\sim750$~GeV.

Yet, another consideration is that $\Phi$ should have a large decay into a pair of $A_1$'s. Naively, the SM fermionic final states would offer a sizable 
competition:
\begin{align}
 &\Gamma[\Phi\to t\bar{t}]\sim\frac{3G_Fm_t^2}{4\sqrt{2}\pi}m_{\Phi}\left(1-4\frac{m_t^2}{m_{\Phi}^2}\right)^{3/2}\left(\frac{g_{\Phi t\bar{t}}}{g^{\rm SM}_{\Phi t\bar{t}}}\right)^2\sim(30~\mbox{GeV})\left(\frac{g_{\Phi t\bar{t}}}{g^{\rm SM}_{\Phi t\bar{t}}}\right)^2\;,\\
 &\Gamma[\Phi\to b\bar{b}]\sim\frac{3G_Fm_b^2}{4\sqrt{2}\pi}m_{\Phi}\left(1-4\frac{m_b^2}{m_{\Phi}^2}\right)^{3/2}\left(\frac{g_{\Phi b\bar{b}}}{g^{\rm SM}_{\Phi b\bar{b}}}\right)^2\sim(0.03~\mbox{GeV})\left(\frac{g_{\Phi b\bar{b}}}{g^{\rm SM}_{\Phi b\bar{b}}}\right)^2\;,
\end{align}
with $m_{\Phi}\simeq750$~GeV and the relative fermionic couplings to their SM counterparts $\frac{g_{\Phi t\bar{t}}}{g^{\rm SM}_{\Phi t\bar{t}}}$ and $\frac{g_{\Phi b\bar{b}}}{g^{\rm SM}_{\Phi b\bar{b}}}$. On the other hand, 
the decay width into $2A_1$ can be estimated as (still for $m_{\Phi}\sim750$~GeV):
\begin{equation}\label{HAAdecay}
 \Gamma[\Phi\taa]=\frac{g^2_{\Phi A_1A_1}}{32\pi m_{\Phi}}\sqrt{1-4\frac{m_{A_1}^2}{m^2_{\Phi}}}\simeq(1\cdot10^{-5}~\mbox{GeV}^{-1})\,g^2_{\Phi A_1A_1}~.
\end{equation}
For the heavy doublet state, the fermionic couplings are approximately determined by $\tan\beta$: $\frac{g_{H_Dt\bar{t}}}{g^{\rm SM}_{H_D t\bar{t}}}\simeq\tan^{-1}\beta$ and 
$\frac{g_{H_Db\bar{b}}}{g^{\rm SM}_{H_Db\bar{b}}}\simeq\tan\beta$. The fermionic channels will thus typically give a width of order $1$~GeV at least, for moderate $\tan\beta=\mathcal{O}(10)$. On the other hand, 
the leading terms in the couplings of a CP-even doublet with a pair of CP-odd singlets read $g_{H_DA_1A_1}\sim\sqrt{2}\lambda(\lambda+\kappa)\cos{2\beta}\,v$. Considering 
$\lambda,\kappa =\mathcal{O}(0.1)$ -- in any case, $\lambda^2+\kappa^2\lsim0.5$ if we want the model to 
remain perturbative up to the GUT scale, we observe that $\Gamma[H_D\taa]\ll \mathcal{O}(\mbox{GeV})$, so that the associated branching ratio is suppressed
in view of the fermionic channels. On the other hand, for $H_S$, the decay channels into SM particles are naturally suppressed while the decay into light CP-odd singlets
is large: $\Gamma[H_S\taa]$ can be read from Eq.~\eqref{HAAdecay} with the replacement $g_{\Phi A_1A_1}\leftarrow g_{H_SA_1A_1}\sim\sqrt{2}\kappa\left(
2 \frac{\kappa}{\lambda}\mu-A_{\kappa}\right)$. If this state $H_S$ were at ${\sim}750$~GeV, then $2 \frac{\kappa}{\lambda}\mu\simeq750$~GeV, while $A_{\kappa}$ is
negligible (from the low mass of the CP-odd singlet). We thus obtain that $\Gamma[H_S\taa]$ is of order $1$~GeV, as long as $\kappa\gsim0.25$. Therefore, 
the branching ratio\footnote{It should be noted that the other singlet-like channel $H_S\to\tilde{\chi}_s\tilde{\chi}_s$, $\tilde{\chi}_s$ denoting the singlino state, is kinematically closed, as
the singlino mass is also of order $2 \frac{\kappa}{\lambda}\mu\simeq750$~GeV.} $\Phi\to\Sigma\Sigma$ pleads for the identification $\Phi\sim H_S$, hence $2 \frac{\kappa}{\lambda}\mu\simeq750$~GeV.

These two apparently conflicting requirements, $\Phi\sim H_D$ for the production and $\Phi\sim H_S$ for the decay, can actually be reconciled if one keeps
in mind that the mass states $H_2$ and $H_3$ are in fact admixtures
(essentially) of $H_D$ and $H_S$. Provided the mixing is large, then
the mass states will combine the properties of both their doublet and
singlet CP-even components. Two interpretations are then possible for
$\Phi$: 
\begin{itemize}
 \item The first is that the diphoton excess corresponds to only one of the two states $H_2$ or $H_3$, while the other would give a subdominant effect due to the 
details of the singlet--doublet mixing. Considering the somewhat large width of order ${\cal O}(45~\mbox{GeV})$ which seems associated with the excess, this interpretation
tends to imply a very large $\Gamma[H_S\taa]={\cal O}(45~\mbox{GeV})$, which could only be achieved with $\kappa\geq1$, that is outside the perturbative regime. 
We will not discuss this scenario any further.%
\footnote{While our study was already in progress, we became aware of the discussion in \cite{UETalk}.}
 \item The second possibility is that both states are sufficiently close in mass to appear as a single excess. The very large binning makes such a scenario
plausible, even for mass differences of order ${\cal O}(20$--$40~\mbox{GeV})$. Then, whatever the width of the physical states $H_2$ and $H_3$, the associated signal would
look like a broad resonance; this will be discussed in
  Section~\ref{sec:diphoton-today}. From previous considerations, we see
that the actual widths of $H_2$ and $H_3$ in the considered regime would
be of order  
${\cal O}(1~\mbox{GeV})$. Their minimal mass difference (for a maximal mixing) can be inferred from the off-diagonal element of the squared-mass matrix Eq.~\eqref{M2S}:
\begin{equation}\label{massgap}
 \left.m_{H_3}-m_{H_2}\right|_{\mbox{\tiny min}}\simeq\frac{\kappa v}{2}\left|\sin{4\beta}\right|\;,
\end{equation}
where we have used $M_A\simeq750~\mbox{GeV}\simeq2\frac{\kappa}{\lambda}\mu$. For $\kappa={\cal O}(0.3)$ and $\tan\beta={\cal O}(10)$, we obtain a typical spread of $10$~GeV in 
mass. This is the scenario we will be focusing on in the following.
\end{itemize}
From now on, we thus assume $M_A\simeq750~\mbox{GeV}\simeq2\frac{\kappa}{\lambda}\mu$, which causes a strong mixing between $H_D$ and $H_S$, so that
$H_{2,3}\sim\frac{1}{\sqrt{2}}\left[H_D\pm H_S\right]$, both states having masses close to $750$~GeV.


\subsection{Other phenomenological constraints}

After identifying how the NMSSM Higgs spectrum could provide an
interpretation of the diphoton excess via the $pp\to
\Phi\to2(\Sigma\to\gamma\gamma)$ topology, we will consider  
additional constraints and consequences on this scenario. 

The SM-like properties of the Higgs state at ${\sim}125$~GeV place a major phenomenological limit on the existence of a light pseudoscalar: as a general
rule, the channel $H_{\rm SM}\taa$ should remain subdominant -- compared to the SM width $\Gamma_{\rm SM}\simeq4$~MeV -- or it would induce a sizable unconventional 
decay of the state at ${\sim}125$~GeV, which would cause tensions with the
results of the LHC Run-I. In our case, with a very light $A_1$ being experimentally indistinguishable
from a photon, the limits on $H_{\rm SM}\taa$ are even more severe in order to avoid a large apparent decay $H_{\rm SM}\to\gamma\gamma$, again in contradiction with 
the Run-I results. The corresponding width can be estimated as
\begin{equation}
 \Gamma[H_{\rm SM}\taa]=\frac{g^2_{H_{\rm SM}A_1A_1}}{32\pi m_{H_{\rm SM}}}\sqrt{1-4\frac{m_{A_1}^2}{m^2_{H_{\rm SM}}}}\simeq(8\cdot10^{-5}~\mbox{GeV}^{-1})\,g^2_{H_{\rm SM}A_1A_1}\;.
\end{equation}
The condition $\Gamma[H_{\rm SM}\taa]\ll \Gamma_{\rm SM}\cdot \BR^{\rm SM}[H_{\rm SM}\tgg]\sim8\cdot10^{-6}$~GeV translates into $g_{H_{\rm SM}A_1A_1}<0.3$~GeV. 
The tree-level coupling $g_{H_{\rm SM}A_1A_1}$ reads
\begin{multline}\label{HAAcoup}
 \frac{g_{H_{\rm SM}A_1A_1}}{\sqrt{2}}\simeq-\kappa A_{\kappa}\,S_{13}(1-P_d^2)+\left[\lambda^2\cos^2\beta-\frac{1}{2}\left(\frac{M_Z}{v}\right)^2\cos^2{2\beta}\right]
 v\,\sqrt{1-S_{13}^2}P_d^2\\
 +\lambda\mu\left[1+\frac{\kappa}{2\lambda}\sin{2\beta}\left(1+\frac{\lambda}{2\kappa}\frac{M_A^2}{\mu^2}\sin{2\beta}\right)\right]S_{13}P_d^2-2\lambda \kappa v\,S_{13}P_d\sqrt{1-P_d^2}\\
 +\lambda \left(\lambda+\kappa\sin{2\beta}\right)v\,\sqrt{1-S_{13}^2}(1-P_d^2)-3\kappa\mu\left[1-\frac{\lambda}{6\kappa}\frac{M_A^2}{\mu^2}\sin{2\beta}\right]\sqrt{1-S_{13}^2}\sqrt{1-P_d^2}P_d\;.
\end{multline}
One should observe that, typically, $\lambda,\kappa ={\cal O}(0.1)$, 
$2<\tan\beta\lsim {\cal O}(10$--$20)$, and $S_{13},P_d\ll1$. As a result,
the two `dangerous' terms in 
Eq.~\eqref{HAAcoup} are those of the last line. Considering Eq.~\eqref{Pd} and $S_{13},P_d\ll1$, the condition on $g_{H_{\rm SM}A_1A_1}$ implies
\begin{equation}
\lambda^2v\left|1+\frac{\kappa}{\lambda}\sin{2\beta}-\frac{9}{4}\left(\frac{2\kappa\mu}{\lambda M_A}\right)^2\left[1-\frac{\lambda}{6\kappa}\frac{M_A^2}{\mu^2}\sin{2\beta}\right]\right|<0.2~\mbox{GeV}
\ \ \ \Rightarrow\ \ \ \lambda^2\left[1-\frac{2\kappa}{\lambda}\sin{2\beta}\right]\lsim1\cdot10^{-3}\;,
\end{equation}
where we have applied the mass condition $M_A\simeq750~\mbox{GeV}\simeq2\frac{\kappa}{\lambda}\mu$ for the final step.

A first alternative would thus consist in 
choosing $\lambda\lsim3\cdot10^{-2}$, which will also lead to $\kappa\lsim0.1$: then, however, the decay $H_S\taa$ no longer competes with the fermionic
decays of $H_D$ and the mixing among $H_S$ and $H_D$ is reduced. Our scenario would thus be invalidated. We will thus prefer to keep $\lambda\gsim0.1$ and satisfy 
the limit from $\Gamma[H_{\rm SM}\taa]$ with the condition 
$\frac{\kappa}{\lambda}\sin{2\beta}\sim0.5$. It should be noted that this condition is actually less ad hoc than it looks at first glance: indeed, given the mass condition 
$M_A\simeq750~\mbox{GeV}\simeq2\frac{\kappa}{\lambda}\mu$, $A_{\lambda}\simeq-\frac{M_A}{2}\left[1-\frac{2\kappa}{\lambda}\sin{2\beta}\right]$, so that, together 
with the requirement of a small $A_{\kappa}$, the assumption $\frac{\kappa}{\lambda}\sin{2\beta}\sim0.5$ places us naturally in the $R$-symmetry limit of the NMSSM.

On the other hand, the perturbativity of the couplings up to the GUT scale approximately implies $\lambda^2+\kappa^2\lsim0.5$. Using the condition $\frac{\kappa}{\lambda}\sin{2\beta}\sim0.5$
then places an upper bound on $\lambda\lsim\sin{2\beta}\sqrt{\frac{2}{1+4\sin^2{2\beta}}}$. Moreover, our scenario requires that the width $\Gamma[H_S\taa]$
remains competitive in view of the fermionic decays of $H_D$, $\Gamma[H_D\to t\bar{t}/b\bar{b}]$: considering the production cross sections of $H_D$, the efficient 
branching ratio $\BR[A_1\to\gamma\gamma]\simeq0.99$, mediated by the pion, and the magnitude of the diphoton excess at ATLAS and CMS, the condition on $\kappa$ can be lowered 
to $\kappa\gsim0.1$ -- instead of $0.25$ as we discussed above -- and translates into a lower bound of $\lambda\sim2\kappa\sin{2\beta}\gsim0.2\sin{2\beta}$.

A further implication of $\frac{\kappa}{\lambda}\sin{2\beta}\sim0.5$ together with the mass condition $M_A\simeq750~\mbox{GeV}\simeq2\frac{\kappa}{\lambda}\mu$
reads $\mu\sim M_A\sin{2\beta}$. An immediate consequence is that $\mu\leq 750$~GeV, and even $\mu\leq 375$~GeV as soon as $\tan\beta\gsim3.7$, so that the 
higgsino states will typically be light, i.e.\ for $\tan\beta\gsim3.7$
$H_2$ and $H_3$ can have a non-negligible decay rate to higgsinos.
This will dilute 
somewhat more the $H_{2,3}\taa$ rates, although the typical widths are of order ${\sim}(15~\mbox{GeV})\lambda^2
\left(1-\frac{\lambda^2}{\kappa^2}\right)^{3/2}$. While nothing forbids that the gauginos also intervene at a low mass, we will assume in the following 
that they are heavier. A neutral higgsino is thus the lightest supersymmetric particle (LSP), which is consistent with cosmological limits on the dark matter 
density: the thermal higgsino relics are typically below the observed value of the relic density~\cite{Drees:2004jm}. However, scenarios with low thermal relic density can be consistent with the measured abundance~\cite{Baer:2010kd,Profumo:2003hq,Fujii:2001xp}.

However, $\mu$ cannot be too small, since this would contradict the unsuccessful results of LEP in chargino searches \cite{LEPchargino}. Using the 
limit $\mu\gsim100$~GeV then provides a bound on $\tan\beta$ in which the choice $\frac{\kappa}{\lambda}\sin{2\beta}\sim0.5$ can be conciliated with the mass
requirement $M_A\simeq750~\mbox{GeV}\simeq2\frac{\kappa}{\lambda}\mu$: $\tan\beta\lsim15$.

In \cite{Ellwanger:2016qax}, limits from $\Upsilon$ decays play a determining
role in constraining the parameter space: the Wilson formula 
\cite{Wilczek:1977pj} gives $\BR[\Upsilon(1S)\to\gamma A_1]\sim1\cdot10^{-4}(P_d\,\tan\beta)^2$, which would typically lead to a ${\sim}10^{-6}$--$10^{-5}$ effect in 
$\Upsilon(1S)$ radiative decays (and similar values for $\Upsilon(2,3S)$). However, we are not aware of experimental limits applying to a pseudoscalar with mass close
to that of the pion. Searches in radiative $\Upsilon$ decays typically ignore mass scales below $0.2$~GeV, except for invisible final states, which are not relevant
in our scenario. Similar contributions of the $A_1$ to the radiative decays of $J/\psi$ are suppressed, due to the smaller charm mass and the $\tan^{-2}\beta$ 
suppression of the $A_1$-charm coupling. Furthermore, as in~\cite{Ellwanger:2016qax}, we find that the impact of $A_1$ in radiative $Z$-decays is orders of
magnitude below the experimental bounds~\cite{PDG,Jaeckel:2015jla}.

As \cite{Andreas:2010ms} pointed out, the presence of a light Higgs pseudoscalar generically leads to tensions in flavor physics. Limits from
invisible decays do not apply in our scenario, as $A_1$ would decay within ${\sim}1$~cm 
at $B$-factories, but the rare transitions $B \to K e^+ e^-$ and $K \to \pi e^+ e^-$ should be considered carefully. Indeed, following \cite{Hiller:2004ii}, one observes that such 
transitions can be mediated by a light $A_1$, as $\BR[A_1\to e^+e^-]$ is
sizable (at the percent level) below the $\mu^+\mu^-$ threshold. The strongest limit comes from \cite{Batley:2009aa}:
\begin{equation}\label{eq:kdecay}
 \mathrm{\BR}^{\mathrm{NA48/2}}(K^\pm \to \pi^\pm e^+ e^-) = (3.11 \pm 0.12) \times 10^{-7}\,.
\end{equation}
The actual bound is in fact much stronger than one can infer from the decay rate alone. The $e^+ e^-$ spectrum is well measured \cite{Batley:2009aa}, with low background and 
small theory uncertainty: therefore a peak in the $e^+e^-$ invariant mass spectrum would be clearly visible. This implies that the $A_1$ contribution has to be strongly 
suppressed: even though, in our configuration, $\BR[A_1\to e^+e^-]$ falls below the percent level due to the large pion-mediated width, the typical magnitude of the effective 
$\bar{b}sA_1$ and $\bar{s}dA_1$ couplings, $C_A\sim10^2$--$10^3~\mbox{GeV}^2$ and $C_A'\sim10^0$--$10^1~\mbox{GeV}^2$, would result in an excess -- see e.g.\ Eqs.~(21) and (25) in
\cite{Andreas:2010ms}. The conclusion that these flavor-changing
processes exclude the considered scenario would be premature,
however. First, one should keep in mind that such 
$A_1$-mediated signals may hide in the background of the pion, due to the proximity in mass.
Then the actual size of $C_A$ and $C_A'$ depends on the details of the sfermion spectrum and, in particular, these coefficients vanish in the super-GIM limit~\cite{Hiller:2004ii}. The sfermion sector is largely free in what precedes our analysis: its only role so far was to ensure the correct mass for the SM-like state via 
radiative corrections -- this essentially translates into a scalar top spectrum of a few TeV, or very large mixing in the stop sector. We check that $C_A$ and $C_A'$ can be made arbitrarily small
for suitable choices of squark spectra, so that flavor constraints --
and not only those involving a $A_1\to e^+e^-$ decay -- can be generally
circumvented; see benchmark point {\bf P2} below.

More general limits on the spectra, such as $B\to X_s\gamma$ or $B_s\to
\mu^+\mu^-$, should 
also be considered~\cite{Domingo:2015wyn}, but note that the already large $m_{H^{\pm}}\simeq750$~GeV, the moderate value of $\tan\beta$, the flexibility of
the squark spectra, and the fact that $A_1$ is off-resonance contrive to place our scenario within $95\%$ of these flavor constraints.

In Ref.~\cite{Andreas:2010ms} it was argued that important constraints on
Higgs-like pseudoscalars can be obtained in beam dump experiments. The most
sensitive one is the CHARM search for axions~\cite{Bergsma:1985qz}. Using
Eq.~(3) of Ref.~\cite{Bergsma:1985qz} one can estimate the $F_X$ parameter to
be $F_X < 10$~GeV, assuming that $\Gamma[A_1 \to \gamma \gamma]/\Gamma[ \pi^0 \to
  \gamma \gamma] > 10^{-4}$ as required by the decay length $\lsim 0.5$~m of $A_1$
at the LHC. Looking at the exclusion plot in Fig.~4 of~\cite{Bergsma:1985qz}
one immediately sees that this is way below sensitivity of the experiment. In
any case, with this decay length at the LHC, the decay length at CHARM would
be of order $\mathcal{O}(1)$~cm. After 60~cm the flux suppression would be ${\sim}
2^{60} \simeq 10^{18}$. Taking into account that the CHARM detector was 480~m away
from the target and that initial $A_1$ flux was $< 10^{17}$ one clearly sees
that possibly no pseudoscalars could have reached the detector. 

Finally, the anomalous magnetic moment of the muon may also be of relevance: a light $A_1$ is indeed known to widen the discrepancy between the prediction
of the model and the experimental measurement; see \cite{Domingo:2008bb}. Yet, the moderate value of $\tan\beta$ and the presence of light higgsinos 
concur to make the supersymmetric corrections to $(g-2)_{\mu}$ the dominant new-physics effect. The overall contribution thus improves the agreement with the BNL 
measurement as compared to the SM. Placing $(g-2)_{\mu}$ within $2\,\sigma$ of the experiment remains problematic, however, and can be achieved only in the upper
reach of $\tan\beta\sim15$. It should be noted that this observable depends on the details of the slepton masses, such that lighter smuons and sneutrinos would improve 
the situation. The LHC searches on light neutralinos and sleptons will be discussed in the following section.

\subsection{Favored parameter space\label{favpar}}
To summarize this analysis, it appears that most of the parameters in the NMSSM Higgs sector are fixed or bounded in the scenario that we consider:
\begin{itemize}
 \item $M_A\simeq750$~GeV enables a sizable production of the state(s) at ${\sim}750$~GeV via a significant $H_ D$ component;
 \item $\kappa\simeq\frac{\lambda}{2\sin{2\beta}}$ ensures a suppressed decay $H_{\rm SM}\taa$; furthermore, $\kappa\gsim0.1$ allows for a competitive $\Gamma[H_S\taa]$ 
as compared to the fermionic decays of $H_D$; finally, $\kappa$ determines the separation in mass for the states at ${\sim}750$~GeV;
 \item $\mu\sim M_A\sin{2\beta}$ is fixed both by the requirement $2 \frac{\kappa}{\lambda}\mu\simeq750$~GeV, conditioning the presence of a singlet-like component at 
${\sim}750$~GeV, with the significant decay to pseudoscalars, and by the condition on $H_{\rm SM}\taa$;
 \item $\lambda$ is bounded as $\frac{0.4\tan\beta}{1+\tan^2\beta}\lsim\lambda\lsim\frac{2\sqrt{2}\tan{\beta}}{\sqrt{1+18\tan^2{\beta}+\tan^4{\beta}}}$: this results from the conditions of a suppressed 
decay $H_{\rm SM}\taa$, which would spoil the interpretation of the LHC Run-I results, of perturbativity up to the GUT scale and of a sizable $\Gamma[H_S\taa]$;
moreover, the light CP-odd Higgs would be long lived if $\lambda$ were too small;
 \item $\tan\beta\lsim15$ is constrained by the lower bound on chargino searches $\mu\gsim100$~GeV, as a result of the various correlations; note that 
$\tan\beta={\cal O}(10)$ satisfies the requirements on the fermionic decays of the states at ${\sim}750$~GeV -- which should remain moderate;
 \item $A_{\kappa}\lsim {\cal O}(0.1) \gev$ conditions a light CP-odd singlet; note that, together with the requirement $A_{\lambda}\to0$, which, in our scenario, follows
the assumptions on $\kappa$, $\lambda$, $\mu$ and $M_ A$, $A_{\kappa}\to0$ places us in the approximate $R$-symmetry limit of the NMSSM, and that $A_1$ thus appears as the
pseudo-Goldstone boson of this $R$-symmetry.
\end{itemize}
Moreover, the requirements of a ${\sim}125$~GeV mass for the SM-like Higgs state and flavor physics constrain the squark spectra, while $(g-2)_{\mu}$ and 
slepton searches impact the slepton spectrum. We stress that the singlino and higgsino masses are essentially determined by the choices in the Higgs sector and that 
light higgsinos (constituting the LSP in the simplest
configuration) appear as a 
trademark of this scenario. 

It should be noted that most of the properties of the Higgs sector
can be transposed 
to a simpler singlet extension of the 2HDM, without the complications in the supersymmetric spectrum and with increased number of parameters and degrees of freedom. 
In the latter setup, one should consider the low-energy constraints more carefully, however, as charged-Higgs contributions to $B\to X_s\gamma$ 
can no longer be balanced by the SUSY loops and the anomalous magnetic moment of the muon would suffer from the negative contributions driven by the loop
involving the muon and the light pseudoscalar if the 2HDM is of Type II (which determines the
production of the states at ${\sim}750$~GeV). It should be kept in
  mind that the singlet $+$ 2HDM framework receives no deep  
motivation from the hierarchy problem, DM or gauge unification.

Naturally, certain attractive features of the NMSSM Higgs sector, such as the possibility of a light CP-even singlet, appear as a necessary sacrifice in 
order to conciliate an interpretation of the ${\sim}750$~GeV excess with
the parameter space and constraints of the NMSSM. Moreover, it could be argued
that the mechanisms which we invoke -- from the sizable singlet--doublet
mixing at ${\sim}750$~GeV, or the condition of a $A_1$--$\pi^0$ interplay, 
to the collimated diphoton decays, indistinguishable from a single
photon -- are quite elaborate. Still, it is remarkable that all 
the necessary properties to fit the signal can be united in a phenomenologically realistic way within as theoretically simple a model as the NMSSM, without 
e.g.\ requiring additional ad hoc matter.


\subsection{Benchmark points}
To investigate the NMSSM parameter space more thoroughly than the derivation at tree level allows, and account for e.g.\ higher-order corrections or 
verify various phenomenological constraints, we use the public package \verb|NMSSMTools 4.8.2| \cite{NMSSMTools}. The Higgs spectrum is computed with precision setting $2$, 
i.e.\ including full one-loop, Yukawa-driven two-loop as well as pole corrections \cite{Degrassi:2009yq}. Note that we dismiss the width and branching fractions
computed by this code for the light $A_1$ as they do not implement the effect of hadronic states. We simply tune the mass $m_{A_1}$ to ${\sim}135$~MeV and then invoke a mixing 
with the pion at the level of $\theta\sim10^{-2}$ -- in practice, this might require further adjustment in the choice of $m_{A_1}$ but this can be achieved with 
completely negligible consequences for the rest of the spectrum. 
Additionally, \verb|NMSSMTools| is interfaced \cite{Domingo:2015eea} with \verb|HiggsBounds 4.2.1| \cite{HiggsBounds} and \verb|HiggsSignals 1.4.0|
\cite{HiggsSignals} in order to test the properties of the Higgs sector in view of current collider limits. However, we still have to check `by hand' that 
the decay $H_ 1\taa$ does not induce a large apparent $H_ 1\tgg$ branching ratio. The sparticle decays are obtained with \verb|NMSDecay| \cite{NMSDecay} 
and the Higgs production cross sections at the LHC are obtained with
\verb|SusHi 1.5.0| \cite{SusHi}, interfaced with \verb|LHAPDF 5.9.1| \cite{LHAPDF} and using MSTW 
parton distribution functions (PDFs) at NNLO \cite{MSTW}. At the outcome of this search, phenomenologically realistic points 
exhibiting the characteristics which we described above are obtained and presented in Table~\ref{points}.

\begin{table*}[tbh!]

\begin{center}
\scalebox{0.9}{
\null\hspace{-1cm}\begin{tabular}{|c|c|c|c|c|c|c|c|c|c|}
\hline
                                       & {\bf P1}  & {\bf P2}  & {\bf P3}  & {\bf P4}   & {\bf P5}  & {\bf P6} & {\bf P7}  & {\bf P8}  & {\bf P9}  \\\hline\hline
\multicolumn{10}{|c|}{Parameters} \\\hline\hline
$\lambda$                             & 0.1       & 0.08      & 0.08      & 0.06       & 0.15      & 0.21     & 0.2       & 0.13      & 0.05      \\ \hline
$\kappa$                              & 0.25      & 0.2       & 0.24      & 0.22       & 0.19      & 0.265    & 0.2       & 0.26      & 0.17      \\ \hline
$\tan\beta$                           & 10        & 10        & 12        & 15         & 5         & 5        & 4         & 8         & 14        \\ \hline
$\mu$ (GeV)                           & 150       & 150       & 127       & 103        & 296       & 296.5    & 375       & 188       & 110       \\ \hline
$M_A$ (GeV)                           & 760       & 784       & 780       & 775        & 785.5     & 785.5    & 810       & 770       & 765       \\ \hline
$A_{\kappa}$ (GeV)                    & 0.003059  & 0.0573065 & 0.0151443 & 0.0012258  & 0.149903  & 0.303953 & 0.4206824 & 0.025274  & $-$0.0017404\\ \hline
$m_{\tilde{Q}}$ (TeV)                 & 1.75      & 10        & 3         & 3          & 10        & 10       & 15        & 3         & 2         \\ \hline
$A_{t}$ (TeV)                         & $-$4      &$-$8.519135& $-$5      & $-$5       & $-$16     & $-$14    & $-$35     & $-$6      & $-$4        \\ \hline
$m_{\tilde{L}}$ (TeV)                 & 0.3       & 0.3       & 0.3       & 0.3        & 0.305     & 0.32     & 0.4       & 0.4       & 0.4       \\ \hline
$M_2$ (TeV)                           & 1         & 1         & 2         & 1          & 1         & 1        & 1         & 1         & 1         \\ \hline\hline
\multicolumn{10}{|c|}{Higgs spectrum}  \\ \hline\hline
$m_{H_1}$ (GeV)                       & 124       & 125       & 125       & 125        & 125       & 124      & 125       & 125       & 125       \\ \hline
$m_{H_2}$ (GeV)                       & 741       & 740       & 753       & 748        & 734       & 726      & 733       & 738       & 744       \\ \hline
$m_{H_3}$ (GeV)                       & 758       & 754       & 766       & 758        & 757       & 759      & 763       & 760       & 753       \\ \hline
$m_{A_1}$ (GeV)                       & 0.135     & 0.135     & 0.135     & 0.135      & 0.135     & 0.135    & 0.135     & 0.135     & 0.135     \\ \hline
$m_{A_2}$ (GeV)                       & 750       & 747       & 759       & 752        & 744       & 744      & 750       & 749       & 750       \\ \hline
$m_{H^{\pm}}$ (GeV)                   & 754       & 751       & 763       & 757        & 747       & 746      & 753       & 753       & 754       \\ \hline
\multicolumn{10}{|c|}{$A_1$ mixing}    \\ \hline\hline
$P_d$                                 & 0.023     & 0.019     & 0.018     & 0.014      & 0.036     & 0.050    & 0.047     & 0.031     & 0.012     \\ \hline\hline
\multicolumn{10}{|c|}{Higgsinos}  \\ \hline\hline
$m_{\tilde{\chi}_1^0}$ (GeV)          & 147       & 149       & 124       & 100        & 294       & 294      & 370       & 185       & 107       \\ \hline
$m_{\tilde{\chi}_2^0}$ (GeV)          & 158       & 160       & 135       & 111        & 310       & 311      & 393       & 197       & 117       \\ \hline
$m_{\tilde{\chi}_1^{\pm}}$ (GeV)      & 152       & 155       & 130       & 105        & 303       & 303      & 384       & 191       & 112       \\ \hline
\end{tabular}}
\end{center}
\caption{Benchmark points: NMSSM input and masses; 
we furthermore choose the trilinear Higgs-sbottom, stau couplings as
$A_{b,\tau}=1.5$~TeV, and the gaugino mass parameters as
  $2M_1=M_2=M_3/3$.
}\label{points} 
\end{table*}

In Table~\ref{points}, we provide the input for \verb|NMSSMTools| as well as relevant masses. The squark soft mass parameters are all chosen degenerate as 
$m_{\tilde{Q}}$ (for simplicity). So are also the soft masses of the sleptons, $m_{\tilde{L}}$. We observe that the Higgs mass predictions of \verb|NMSSMTools| 
for very heavy sfermions may not be entirely reliable, as a resummation
of large $\log(\frac{m_{\tilde{Q}}}{m_t})$ may be necessary: such effects
are addressed, e.g.\ in~\cite{Drechsel:2016jdg} but the details of the correspondence between the parameters and the spectrum are of secondary importance for our conclusions. 
Note that 
all the considered points satisfy the phenomenological tests implemented within \verb|NMSSMTools| -- except maybe $(g-2)_{\mu}$ (satisfied for {\bf P4} and 
{\bf P9}) -- and \verb|HiggsBounds| within $2\,\sigma$. We make sure that $\BR[H_1\taa]<1\cdot10^{-4}$. The fit values to the Higgs measurement 
at ${\sim}125$~GeV obtained with \verb|HiggsSignals| are all competitive with the SM -- i.e.\ $\chi^2<\chi^2_{\rm SM}$ or $|\chi^2/\chi^2_{\rm SM}-1|\ll1$.

Concerning the flavor constraints associated to $A_1\to e^+e^-$ or, more
generally, to a mediation by
$A_1$, we stress that the proximity in mass of $A_1$ to the pion would
certainly require a more careful analysis on the experimental side, so that
the current experimental limits are likely not to apply. Yet, for
completeness, we wish to show to which extent $A_1$-mediated contributions
to rare $B$ and $K$ decays can be reduced in our scenario.
We thus undertook the task of tuning the parameters in the sfermion 
sector in order to suppress the effective flavor-changing couplings for point {\bf P2} only: it is quite clear that such a requirement can always be applied independently of the 
properties of the Higgs states. For this point, the trilinear
  Higgs-stop coupling $A_t$ is adjusted in such a way that the effective flavor-changing $A_1$ couplings amount to $C_A\simeq1.3\cdot10^{-6}~\mbox{GeV}^2$ 
and $C'_A\simeq2.6\cdot10^{-8}~\mbox{GeV}^2$. These extremely suppressed numbers come at the price of the $7$-digit precision in the value of $A_t$. We note that to simultaneously maintain $m_{H_1}\simeq125$~GeV a significantly heavier squark sector (compared to {\bf P1}) becomes necessary: this is not unexpected as it is unlikely to combine a maximal stop mixing (which provides a large contribution to $m_{H_1}$) and minimal effective flavor-changing $A_1$ couplings with the sole handle of $A_t$.\footnote{
The low $\tan\beta$ range is typically less sensitive to limits from flavor transitions (as $\tan\beta$ is no longer an enhancement factor). Yet, large squark masses also emerge as a necessity to generate $m_{H_1}\simeq125$~GeV because of the lower tree-level Higgs mass ${\sim} M_Z\cos{2\beta}$; note that tree-level NMSSM effects on the SM-like Higgs mass, using large $\lambda$ or singlet--doublet mixing, are difficult to combine with our scenarios.}
The contribution to $\BR[B^0\to K^0e^+e^-]$ is then at the level of ${\sim}10^{-20}$ and, for $\BR[K^{+}\to \pi^{+}e^+e^-]$ at the level of 
${\sim }10^{-23}$. Such effects are far too small to be measurable experimentally. However, if we consider {\bf P1} for comparison, where the sfermion sector was not tailored to 
accommodate these channels, $C_A\sim130~\mbox{GeV}^2$ and $C'_A\sim3~\mbox{GeV}^2$, contributing to the branching ratios at the level of $10^{-4}$ and $10^{-7}$ respectively,
i.e.\ far beyond existing limits. Of course, the precision that we requested in the suppression of $C_A$ and $C'_A$ for point {\bf P2} is not really 
necessary in view of the $e^+e^-$ channels: $C_A\sim1~\mbox{GeV}^2$ and $C_A'\sim0.1~\mbox{GeV}^2$ would be sufficient and can be achieved with the simpler requirement $A_t\simeq-8.5$~TeV for 
{\bf P2}. Furthermore, experimental cuts would typically require $m_{e^+ e^-} > 140$~MeV~\cite{Batley:2009aa}, so that our scenario with $m_{A_1} \simeq m_{\pi^0}$ is not affected by these limits in general. However, we wish to stress that any bound on flavor transitions mediated by $A_1$ could be circumvented in such a fashion. Therefore, we will pay no further 
attention to these flavor limits on the basis that they strongly depend on the
details of the sfermion sector.

We now discuss the phenomenology of the benchmark points.
$\tan\beta$ ranges from $4$ to $15$ in Table~\ref{points}: this implies a variety of regimes for the points under consideration, as we will see later. In particular,
this determines the value of $\mu$, i.e.\ the higgsino spectrum: the lightest neutralino mass varies between $100$~GeV at $\tan\beta=15$ ({\bf P4}) to $370$~GeV at $\tan\beta=4$
({\bf P7}). $\kappa$ and $\lambda$ are always of order $0.1$ and their ratio also depends on $\tan\beta$ (see previous section). Note that larger values of $\lambda$ are accessible
but tend to result in too efficient a cross section for the diphoton signal, as we shall discuss later. The values of $A_{\kappa}$ appear with a sizable number of digits: this 
corresponds to the precision necessary to keep $m_{A_1}$ within $135\pm0.5$~MeV. $M_A$ falls within $10$ to $60$~GeV of $750$~GeV. The squark masses are chosen, together with
the trilinear coupling $A_t$, so as to generate a mass close to ${\sim}125$~GeV for $H_1$: the values fall in the range $1$--$20$~TeV. We have already commented the number of digits
for $A_t$ in {\bf P2}: we aim to check that flavor--transitions mediated by $A_1$ can be made arbitrarily negligible. The slepton masses are taken between $300$ and $400$~GeV,
depending on the mass of the lightest neutralino: they matter only for
$(g-2)_{\mu}$. Finally, we employ hierarchical gaugino masses,
$2M_1=M_2=M_3/3=1$~TeV and trilinear soft
couplings $A_{b,\tau}=1.5$~TeV: they play essentially no role here.

\begin{table*}[tbh!]
\null\hspace{-2cm}
\begin{center}
\scalebox{0.9}{
\begin{tabular}{|c|c|c|c|c|c|c|c|c|c|}
\hline
                                    & {\bf P1}  & {\bf P2}  & {\bf P3}  & {\bf P4}   & {\bf P5}  & {\bf P6} & {\bf P7}  & {\bf P8}  & {\bf P9}  \\\hline\hline
\multicolumn{10}{|c|}{Higgs decays} \\\hline\hline
$\BR[H_1\taa]$&$7\cdot10^{-6}$&$6\cdot10^{-5}$&$1\cdot10^{-5}$&$9\cdot10^{-6}$&$3\cdot10^{-6}$&$3\cdot10^{-7}$&$8\cdot10^{-6}$&$1\cdot10^{-6}$&$5\cdot10^{-6}$\\ \hline\hline
$\Gamma_{H_2}$ (GeV)                            & 1.60      & 1.53      & 2.04      & 2.71       & 1.30      & 1.53     & 1.29      & 1.37      & 1.41      \\ \hline
$\BR[H_2\taa]$                                   & 0.306     & 0.174     & 0.188     & 0.113      & 0.190     & 0.373    & 0.288     & 0.363     & 0.186     \\ \hline
$\BR[H_2\to b\bar{b}]$                           & 0.332     & 0.397     & 0.439     & 0.527      & 0.117     & 0.087    & 0.056     & 0.269     & 0.599     \\ \hline
$\BR[H_2\to t\bar{t}]$                           & 0.094     & 0.121     & 0.064     & 0.032      & 0.533     & 0.357    & 0.551     & 0.186     & 0.046     \\ \hline
$\BR[H_2\to \tau\bar{\tau}]$                     & 0.048     & 0.058     & 0.064     & 0.077      & 0.017     & 0.013    & 0.008     & 0.039     & 0.087     \\ \hline
$\BR[H_2\to \tilde{h}\tilde{h}]$                 & 0.012     & 0.004     & 0.003     & 0.003      & 0.021     & 0.040    & 0         & 0.027     & 0.002     \\ \hline\hline
$\Gamma_{H_3}$ (GeV)                            & 1.92      & 1.55      & 2.00      & 2.28       & 1.52      & 2.09     & 2.27      & 1.71      & 2.05      \\ \hline
$\BR[H_3\taa]$                                & 0.231     & 0.213     & 0.247     & 0.185      & 0.191     & 0.226    & 0.099     & 0.301     & 0.082     \\ \hline
$\BR[H_3\to b\bar{b}]$                           & 0.279     & 0.292     & 0.327     & 0.427      & 0.073     & 0.062    & 0.043     & 0.182     & 0.608     \\ \hline
$\BR[H_3\to t\bar{t}]$                           & 0.096     & 0.104     & 0.055     & 0.029      & 0.452     & 0.395    & 0.655     & 0.162     & 0.052     \\ \hline
$\BR[H_3\to \tau\bar{\tau}]$                     & 0.041     & 0.043     & 0.048     & 0.062      & 0.011     & 0.009    & 0.006     & 0.027     & 0.089     \\ \hline
$\BR[H_3\to \tilde{h}\tilde{h}]$                 & 0.165     & 0.154     & 0.135     & 0.090      & 0.112     & 0.123    & 0.002     & 0.222     & 0.087     \\ \hline\hline
$\Gamma_{A_2}$ (GeV)                            & 2.40      & 2.37      & 3.02      & 4.19       & 2.18      & 2.30     & 2.83      & 1.80      & 2.99      \\ \hline
$\BR[A_2\to\tau\tau]$                            & 0.065     & 0.065     & 0.075     & 0.084      & 0.018     & 0.016    & 0.009     & 0.055     & 0.102     \\ \hline\hline
\multicolumn{10}{|c|}{Higgs production}                                                                                                                    \\ \hline\hline
$\sigma^{gg{\rm f}}_{\mathrm{8 TeV}}[H_2]$ (fb)       & 0.60      & 0.74      & 0.50      & 0.50       & 3.07      & 2.62     & 3.23      & 1.01      & 0.34      \\ \hline
$\sigma^{bbh}_{\mathrm{8 TeV}}[H_2]$ (fb)             & 3.90      & 4.53      & 5.98      & 9.91       & 1.21      & 1.13     & 0.58      & 2.78      & 6.01      \\ \hline
$\sigma^{gg{\rm f}}_{\mathrm{8 TeV}}[H_3]$ (fb)       & 0.60      & 0.55      & 0.35      & 0.32       & 2.40      & 2.84     & 5.02      & 0.88      & 0.48      \\ \hline
$\sigma^{bbh}_{\mathrm{8 TeV}}[H_3]$ (fb)             & 3.37      & 3.00      & 3.84      & 6.17       & 0.71      & 0.82     & 0.59      & 1.95      & 8.20      \\ \hline\hline
$\sigma^{gg{\rm f}}_{\mathrm{13 TeV}}[H_2]$ (fb)      & 2.62      & 3.25      & 2.21      & 2.15       & 10.36     & 11.52    & 14.33     & 4.47      & 1.49      \\ \hline
$\sigma^{bbh}_{\mathrm{13 TeV}}[H_2]$ (fb)            & 20.70     & 24.08     & 32.14     & 53.05      & 6.35      & 5.89     & 3.03      & 14.72     & 32.05     \\ \hline
$\sigma^{gg{\rm f}}_{\mathrm{13 TeV}}[H_3]$ (fb)      & 2.66      & 2.45      & 1.57      & 1.39       & 10.87     & 12.90    & 22.88     & 3.97      & 2.12      \\ \hline
$\sigma^{bbh}_{\mathrm{13 TeV}}[H_3]$ (fb)            & 18.21     & 16.17     & 20.95     & 34.00      & 3.86      & 4.46     & 3.23      & 10.54     & 44.11     \\ \hline\hline
$\sigma^{gg{\rm f}}_{\mathrm{13 TeV}}[A_2]$ (fb)      & 12.97     & 13.42     & 10.46     & 10.14      & 37.73     & 37.79    & 53.74     & 17.10     & 10.41     \\ \hline
$\sigma^{bbh}_{\mathrm{13 TeV}}[A_2]$ (fb)            & 38.62     & 40.05     & 52.81     & 86.19      & 10.19     & 10.20    & 6.25      & 25.01     & 75.85     \\ \hline\hline
\multicolumn{10}{|c|}{$\gamma\gamma$@$750$~GeV}                                                                                                             \\ \hline\hline
$\sigma^{\mathrm{incl}}_{\mathrm{8 TeV}}$ (fb)  & 2.2       & 1.6       & 2.5       & 2.32       & 1.37      & 2.18     & 1.61      & 2.23      & 1.85       \\ \hline
$\sigma^{\mathrm{incl}}_{\mathrm{13 TeV}}$ (fb) & 11.7      & 8.5       & 13.55     & 12.48      & 5.83      & 10.17    & 7.40      & 11.06     & 9.80     \\ \hline
\end{tabular}}
\end{center}
\caption{Higgs branching fractions and production cross sections.}\label{BRprod}
\end{table*}

Table~\ref{points} also provides the Higgs masses: from the discussion above, it should be clear that the parameters have been chosen so that:
\begin{itemize}
 \item $m_{H_1}\simeq125$~GeV corresponds to the SM-like Higgs boson, identified with the ${\sim}125$~GeV signal of the LHC;
 \item $m_{A_1}\simeq135$~MeV, and the $A_1$ should mix with $\pi^0$;
 \item $m_{H_2,H_3}\simeq750$~GeV; consequently, $m_{A_2}$ and $m_{H^{\pm}}$ also fall close to $750$~GeV.
\end{itemize}
The only quantity deserving discussion at this level is the mass-splitting between $H_2$ and $H_3$: it ranges from ${\sim}10$ to ${\sim}30$~GeV, depending on the values of $\kappa$
and $\tan\beta$. As we aimed at a large singlet--doublet mixing close to $50\%$, this mass gap is essentially determined by Eq.~\eqref{massgap}.

Finally, we indicate the magnitude of the doublet component in $A_1$, $P_d$, which plays a central role for the characteristics of this state. It follows the approximate
rule $P_d=0.232\cdot\lambda$, that one can infer from the tree-level definition, Eq.~\eqref{Pd}, together with the various conditions on $M_A$, $\mu$, $\kappa$ and $\sin{2\beta}$
presented in Section~\ref{favpar}.

In Table~\ref{BRprod}, we indicate several Higgs branching fractions as well as the production cross sections in $gg$f and $bbh$ at the LHC for the heavy states. First,
we check that $\BR[H_1\taa]<1\cdot10^{-4}$, so that no non-SM diphoton decay of $H_1$ conflicts with the LHC Run-I results -- the total width of $H_1$ is always SM-like: 
$\Gamma_{H_1}\simeq4\cdot10^{-3}$~GeV. Concerning the heavy CP-even states, their widths fall typically between $1$ and $2$~GeV and are dominated by the fermionic channels 
-- $b\bar{b}$ and/or $t\bar{t}$ depending on $\tan\beta$. $\BR[H_{2,3}\taa]$ is typically at $10$--$30\%$ and comparable for both states (due to an efficient mixing): 
we observe that larger values ($\gsim50\%$) are accessible for larger $\kappa$ ($\lambda$) but such values tend to overshoot the magnitude of the observed diphoton 
cross section. We also provide the branching ratios to higgsinos, regrouping all the decays to the lightest (next-to-lightest) neutralino and chargino states 
($\tilde{h}\ni\{\tilde{\chi}_1^0,\tilde{\chi}_2^0,\tilde{\chi}_1^{\pm}\}$), as well as to $\tau\bar{\tau}$ -- this will be discussed in Section~\ref{futdir} in connection with
future tests of our scenario.

Regarding the production cross sections of the heavy CP-even Higgs states at the LHC, they are essentially driven by the doublet components of these states. They are shared in 
roughly equal proportions by the $H_{2,3}$ states, as a consequence of the ${\sim}50\%$ mixing. The cross section in $gg$f is suppressed as $\tan^{-2}\beta$ (in agreement with
the $\tan^{-1}\beta$ suppression of the $H_D$-coupling to tops) while
that in $bbh$ varies as $\tan^2\beta$ (in accordance with the
$\tan\beta$ enhancement of the $H_D$-coupling  
to bottoms): summing over both states, the production cross section in
$gg$f at $8$~TeV follows the approximate rule $\sigma^{gg{\rm f}}_{\mathrm{8 TeV}}[H_2+H_3]\simeq(130~\mbox{fb})
\tan^{-2}\beta$; that in $bbh$ $\sigma^{bbh}_{\mathrm{8 TeV}}[H_2+H_3]\simeq(0.07~\mbox{fb})\tan^{2}\beta$; at $13$~TeV,
$\sigma^{gg{\rm f}}_{\mathrm{13 TeV}}[H_2+H_3]\simeq
(600~\mbox{fb})\tan^{-2}\beta$ and $\sigma^{bbh}_{\mathrm{13 TeV}}[H_2+H_3]\simeq(0.4~\mbox{fb})\tan^{2}\beta$. At low $\tan\beta\sim4$--$5$, the $gg$f channel thus dominates, while
the $bbh$ is more efficient at large $\tan\beta\sim10$--$15$. Their sum is
maximal at large $\tan\beta$: ${\sim}17$~fb at $8$~TeV and ${\sim}95$~fb at
$13$~TeV, and it is minimal for 
$\tan\beta\sim6$: ${\sim}6$~fb at $8$~TeV and ${\sim}30$~fb at $13$~TeV. The
enhancement factor between $8$ and $13$~TeV ranges from ${\sim}4.75$ at
$\tan\beta=4$ to $5.65$ at $\tan\beta=15$. Considering that the $8$~TeV data
did not show any significant diphoton excess at ${\sim}750$~GeV, one would
prefer this 
enhancement factor to be as large as possible, so that it avoids tensions with the limits from Run-I.
An enhanced associated production with $b$-quarks, which occurs at
large $\tan\beta$ is thus
slightly preferred since the ratio between the 13 TeV and 8
  TeV production cross section is the largest for $b\bar b$ initial
  states. 
The production cross sections of $A_2$ at $13$~TeV are also given. They
follow coarsely the same patterns as their analogues for $H_2/H_3$,
with a larger $gg$f though.

These production cross sections and branching ratios allow us to derive
the relevant cross section for the $pp\to H_2 (H_3) \to2(A_1\tgg)$ process (we assume 
$\BR[A_1\tgg]\simeq0.99$): this quantity is documented for our points in the last two lines of Table~\ref{BRprod}, at $8$ and $13$~TeV. Depending on the characteristics of
our points, $\sigma^{\mathrm{incl}}_{\mathrm{8 TeV}}\sim1$--$2$~fb and $\sigma^{\mathrm{incl}}_{\mathrm{13 TeV}}\sim5$--$13$~fb, which is the relevant order of magnitude for an 
interpretation of the diphoton excess. These figures shall be analyzed with further detail in the following section.


\section{Collider analysis\label{sec:collider}}

\subsection{Analysing the diphoton signal with current data}
\label{sec:diphoton-today}

\begin{figure}
\begin{center}
 \includegraphics[width=0.6\textwidth]{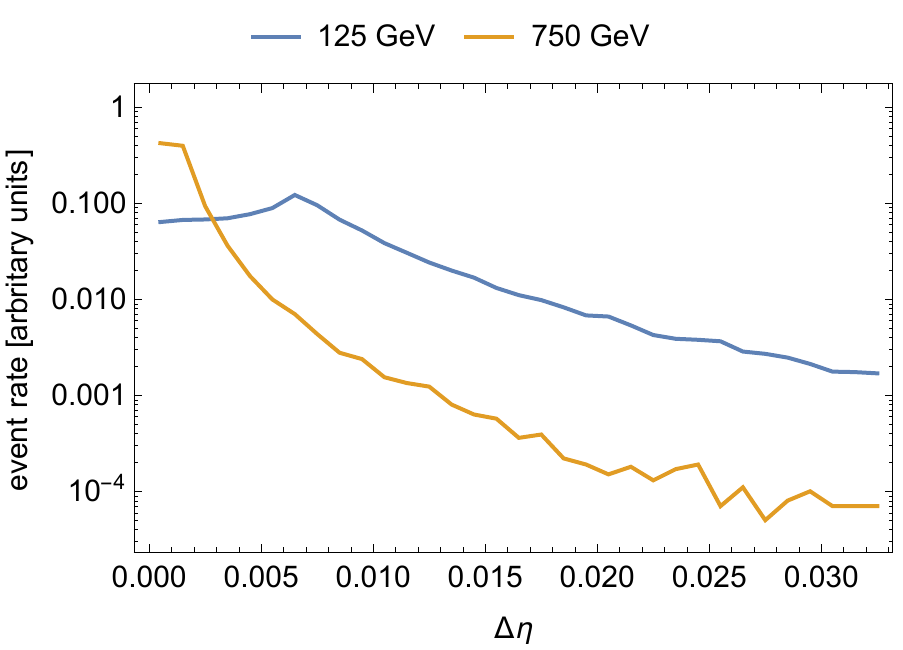}
\end{center}
\caption{The pseudorapidity separation between two photons from the $A_1$ decay, $m_{A_1}=200$~MeV, for the resonance production of the SM-like Higgs boson (dark/blue line) and the hypothetical 750~GeV (light/beige line) scalar. \label{fig:deta}}
\end{figure}

We consider resonant $H_2$ ($H_3$) production,
\begin{equation}
pp\rightarrow H_2\,(H_3)\rightarrow2(A_1\rightarrow \gamma\gamma) + X,\label{eq:signal}
\end{equation}
where $X$ stands for the rest of the event. Due to the large boost of $A_1$, the two photons of the $A_1$ decay will be very collimated and thus the opening angle between both photons in the electronic calorimeter
will be well below the angular resolution of electromagnetic calorimeters
\cite{Aad:2010sp,Khachatryan:2015iwa}. Therefore, our final state resembles a
resonant diphoton final state. The ATLAS conference
note~\cite{ATLASfourphoton2012} studied the signature of a 125~GeV Higgs boson
decaying to four photons at $\sqrt{s}=7$ TeV with
the full data set. The search estimates the efficiency of photon-pair
identification as a single photon at about $85\%$--$95\%$ for photons with
$p_T \approx 100 \gev$ and mass $m_{A_1} = 100$--$200 \mev$. The efficiency heavily
depends on the mass of the heavy resonance, as can be seen in
Figure~\ref{fig:deta} which shows the $\Delta\eta :=
|\eta(\gamma_1)-\eta(\gamma_2)|$ distribution of the two photons from the
$A_1\rightarrow\gamma\gamma$ decay where $\eta$ denotes the
pseudorapidity. The dark (blue) curve shows the result for a Higgs boson
mass of $125 \gev$, while the light (beige) curve for $750 \gev$. As expected
the opening angle of the photons from the 750 GeV resonance is much smaller
compared to the 125 GeV case. However, it is difficult to determine the exact
efficiency without performing a full detector simulation. Hence, we will
choose $\epsilon = 90\%$ in the remainder of the paper. In any case, our
conclusions will not considerably change if we assume a higher efficiency as
the one from the ATLAS study~\cite{ATLASfourphoton2012}. This means that about
$80\%$ of all four-photon events within the fiducial region will be classified
as diphoton events. This choice is further supported by the analysis in
Ref.~\cite{Aparicio:2016iwr}.

In order to test compatibility of the parameter points with experimental results at $\sqrt{s}= 8$~TeV~\cite{Aad:2014ioa,Aad:2015mna,Khachatryan:2015qba,CMS-PAS-EXO-12-045} and 13~TeV~\cite{ATLASdiphoton2015,CMSdiphoton2015} 
we generated parton-level events with {\tt Madgraph 2.3.3}
\cite{Alwall:2014hca} interfaced with the Monte Carlo (MC) generator {\tt Pythia 6.4}
\cite{Sjostrand:2006za} for the parton shower and hadronization. 
We have implemented the 8 and 13 TeV diphoton searches from ATLAS and
CMS \cite{ATLASdiphoton2015,CMSdiphoton2015}  
into the {\tt CheckMATE 1.2.2} framework~\cite{Drees:2013wra} with its
{\tt AnalysisManager} \cite{Kim:2015wza}.  
{\tt CheckMATE 1.2.2} is based on the fast detector simulation {\tt
  Delphes 3.10} \cite{deFavereau:2013fsa} with heavily  
modified detector tunes and it determines the number of expected signal events passing the
selection cuts of the particular analysis. The selection cuts  
for both ATLAS and CMS analyses are shown in Table~\ref{tab:selection13tev}. The resulting signal efficiency varies between 20$\%$ and 60$\%$ depending on the signal region, the experiment and the center-of-mass energy.

\begin{table}
\centering \renewcommand{\arraystretch}{1.3}
\scalebox{1}{
\begin{tabular}{|c|c|}\hline
 ATLAS & CMS \\
\hline
\hline
$p_T(\gamma)\ge$25 GeV  &
$p_T(\gamma)\ge$75 GeV \\ 
\hline
$|\eta(\gamma)|\le2.37$ & $|\eta(\gamma)|\le 1.44$
 or $1.57 \le |\eta(\gamma)| \le 2.5$ 
\\ 
& at least one $\gamma$ with $|\eta(\gamma)|\le1.44$\\
\hline
$E_T^{\gamma_1}/m_{\gamma\gamma}\ge0.4$,
$E_T^{\gamma_2}/m_{\gamma\gamma}\ge0.3$ & $m_{\gamma\gamma}\ge230$ GeV \\ \hline
\end{tabular}}
\caption{Selection cuts of the 13 TeV ATLAS/CMS diphoton searches
  \cite{ATLASdiphoton2015,CMSdiphoton2015}. 
\label{tab:selection13tev} }
\end{table}

Using the above setup we calculate the expected number of
events in the signal regions centered at 750~GeV for each parameter
point {\bf P1}--{\bf P9} in four 8 TeV searches and in two
13 TeV searches. The results are collected in Table~\ref{events}. 
For reference, in column two and three we provide the observed number of
events above the SM background (``sig.'') and the observed $S95$ exclusion
limits calculated using \texttt{CheckMATE}~\cite{Drees:2013wra,Read:2002hq}.
Our benchmark points offer a range of cross sections for the desired
signal, from $5.8$ to $12.7$~fb at 13~TeV, as listed in
  Table~\ref{BRprod}. Points {\bf P1}, {\bf P3}, 
{\bf P4} and {\bf P8}  fit exactly the claimed event rate
from ATLAS~\cite{ATLASdiphoton2015}, but predict too many events in
the CMS signal region \cite{CMSdiphoton2015}; see also the discussion in~\cite{Kim:2015ksf}. The remaining points
fulfill all constraints. In a model independent $\chi^2$ fit we
estimate that the best-fit cross section at 13 TeV using ATLAS and CMS
results is $8.3\fb$~\cite{Kim:2015ksf}, 
with points {\bf P2} and {\bf P7} being closest
in value, cf.\ Table~\ref{BRprod}. A similar analysis for the 8~TeV
data yields $0.5\fb$, while a combination of all the available data
gives a range of cross sections $4.9$--$5.7\fb$ at 13 TeV. The exact
result depends on the details of the production mechanism, but our
benchmark points cover well the desired range. The best fit to all
data is provided by point {\bf P5}.

\begin{table*}[tbh!]
\null\hspace{-2cm}
\begin{center}
\scalebox{0.95}{
\begin{tabular}{|l||c|c||c|c|c|c|c|c|c|c|c|}
\hline
Search                                     & sig.   &  $S95$  & {\bf P1}   & {\bf P2}  & {\bf P3}   & {\bf P4}   & {\bf P5}  & {\bf P6} & {\bf P7}  & {\bf P8}   & {\bf P9}    \\ \hline\hline
\multicolumn{12}{|c|}{$\sqrt{s}=13$~TeV}  \\ \hline\hline
ATLAS13~\cite{ATLASdiphoton2015}           & 16.6   & 27      & 14.1       & 10.3      & 15.4       & 15.1       & 7.1       & 12.3     & 8.9       & 13.4       & 11.8        \\ \hline
CMS13 EBEB~\cite{CMSdiphoton2015}          & 4.5    & 12.8    & $14.4^{*}$ & 10.5      & $15.7^{*}$ & $15.4^{*}$ & 7.2       & 12.6     & 9.2       & $13.7^{*}$ & 12.1        \\ \hline
CMS13 EBEE~\cite{CMSdiphoton2015}          & 0      &  9.5    & 5.4        & 4.0       & 5.9        & 5.8        & 2.7       & 4.7      & 3.4       & 5.1        & 4.6         \\ \hline\hline
\multicolumn{12}{|c|}{$\sqrt{s}=8$~TeV}  \\ \hline\hline
ATLAS8-1407.0653~\cite{Aad:2014ioa}        & 6      & 20      & 15.8       & 11.6      & 16.9       & 16.4       & 9.7       & 15.4     & 11.4      & 15.4       & 13.1        \\ \hline
ATLAS8-1504.05511~\cite{Aad:2015mna}       & 2.6    & 23      & 21.5       & 15.7      & 22.9       & 22.3       & 13.2      & 20.9     & 15.5      & 20.9       & 17.7        \\ \hline
CMS8-EXO-12-045~\cite{CMS-PAS-EXO-12-045}  & 0      & 16      & 9.4        & 6.8       & 10.0       & 9.7        & 5.7       & 9.1      & 6.8       & 9.1        & 7.7         \\ \hline
CMS8-1506.02301~\cite{Khachatryan:2015qba} & 3      & 34      & 16.2       & 11.8      & 17.2       & 16.8       & 9.9       & 15.7     & 11.7      & 15.7       & 13.4        \\ \hline
\end{tabular}}
\end{center}
\caption{Event numbers due to the heavy Higgs production in
  the signal regions of the ATLAS and CMS diphoton searches at
  $\sqrt{s}=8$ and 13~TeV for each of the benchmark scenarios
  considered (we keep two separate signal regions for CMS13). For reference, we provide the observed number of signal events
    above the expected SM background (``sig.''; 0 if the number of expected background events exceeds the number of observed events) and the observed $S95$ exclusion limits
  calculated using \texttt{CheckMATE}~\cite{Kim:2015wza}.
  The event numbers marked with a $^{*}$ 
would be excluded at CL 95\% for the respective channel.
}
\label{events}
\end{table*}

\begin{figure}[htb!]
\begin{center}
\includegraphics[width=0.48\textwidth]{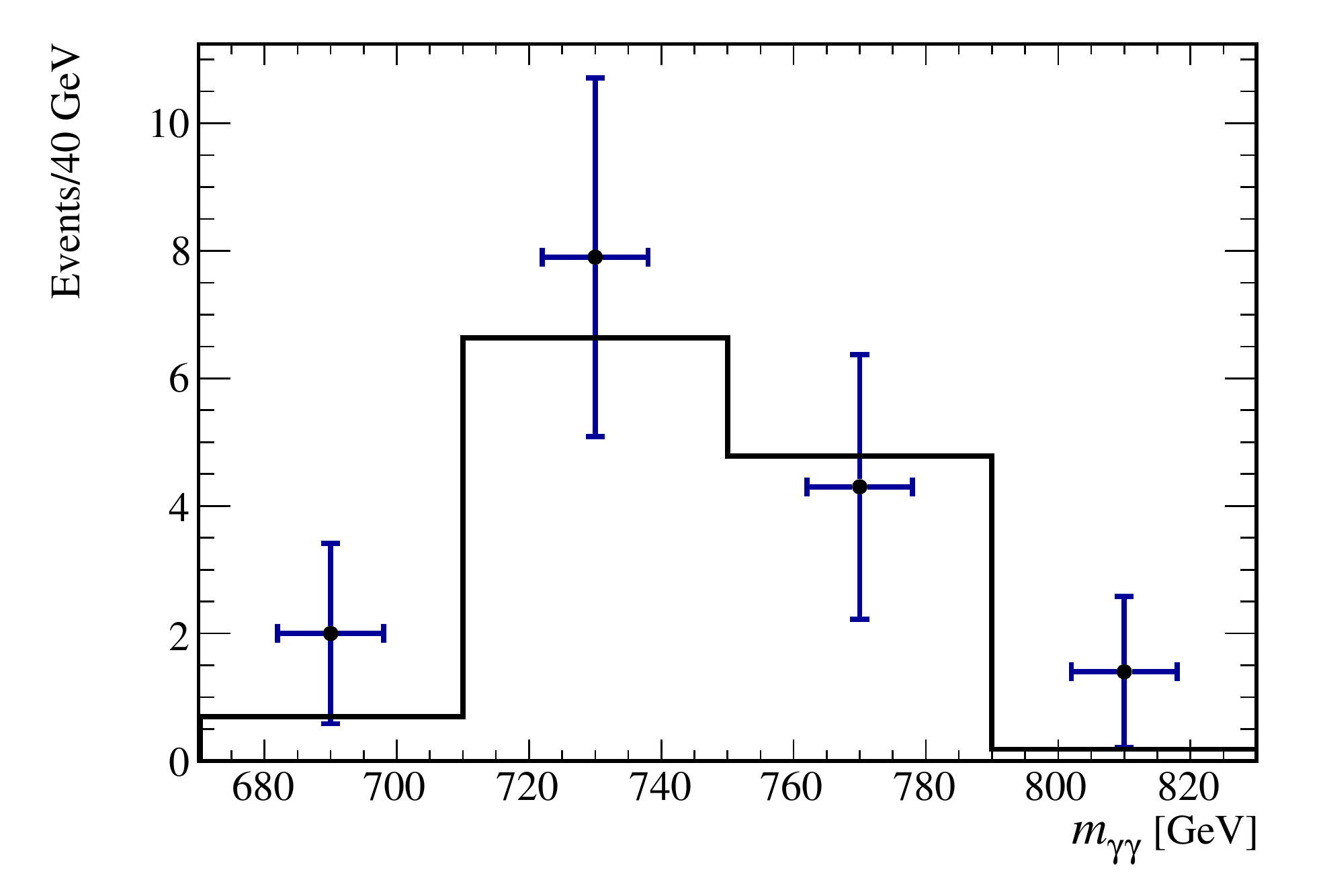} 
\includegraphics[width=0.48\textwidth]{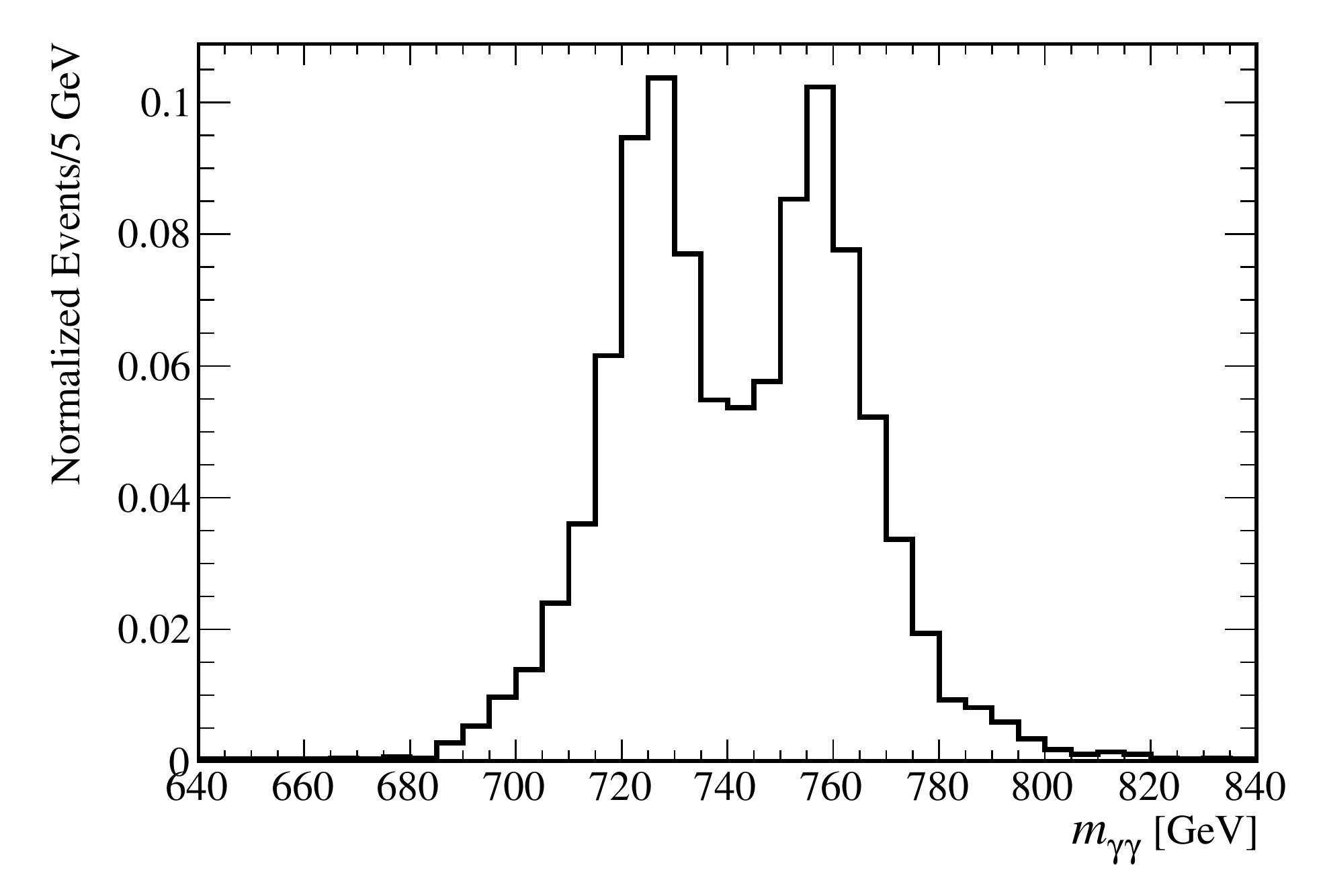}
\caption{Invariant mass distribution of the diphoton resonance of
benchmark point {\bf P6} (black histograms). 
Left: with a bin size of 40~GeV
corresponding to the experimental bin size of the ATLAS search
\cite{ATLASdiphoton2015} and the number of events over background with errors obtained by ATLAS for each point (blue).
Right: with a bin size of 5~GeV showing a twin-peak feature.}
\label{fig:double_resonance}
\end{center}
\end{figure}

In Fig.~\ref{fig:double_resonance}, we show the diphoton invariant mass 
distribution of our diphoton signal for two different bin sizes. We consider
benchmark point {\bf P6} for illustration. The distribution with
the large bin size of 40 GeV corresponds to the experimental bin size of the
ATLAS study 
\cite{ATLASdiphoton2015}, as shown in the left panel.
The experimental photon energy resolution of about
5--10$\%$ would allow for a higher precision \cite{Aad:2014nim} but due to the small statistical
sample, both experiments have to choose a rather large bin size. One can
clearly see that our benchmark point with two scalars cannot be distinguished
from a wide resonance with the current data. For comparison we have
included into this plot the original data from ATLAS after subtracting the expected background. 
One can see that the events predicted for our {\bf P6} benchmark provide a good reproduction of the experimental shape.
We also display in the right panel of Fig.~\ref{fig:double_resonance} the
invariant mass distribution with a $5$~GeV binning. While currently the experimental
resolution in $m_{\gamma\gamma}$ exceeds 10~GeV, one can speculate that further improvements during the current LHC run will be made. 
With the accuracy of ${\sim }5$~GeV and an increased luminosity, the broad
excess, provided it is real, 
might be resolved as two narrow resonances
\cite{Cao:2016cok}. 

\medskip\smallskip
Our estimate of the number of signal events in Table~\ref{events} has a
theoretical uncertainty. The choice of the parton distribution function,
missing higher-order calculations and the details of the parton shower as well
as the fragmentation induce an uncertainty on the signal rate. However, our
final-state configuration is relatively simple and thus the details of the MC
tuning will not significantly alter our results. The size of the uncertainty
from the PDFs can be sizable. The variation
between the different PDFs changes the hadronic cross section. 
In addition, the scale dependence of the production cross section on the renormalization
and factorization scale affects the signal rate. We estimate the sum of  these effects to 
be of the order of 10$\%$~\cite{Heinemeyer:2013tqa}. Furthermore, the
normalization can heavily depend on the value of the bottom Yukawa coupling
\cite{Heinemeyer:2013tqa}. Finally, we did not model the detector
response of 
identifying a photon from the pseudoscalar decay into the diphoton state but
rather choose a flat efficiency factor of $\epsilon=90\%$. Here, one can
assume a conservative uncertainty of about $20\%$ on the signal rate due to
the uncertainty in the photon identification. We conclude that the total
uncertainty is of the order of $\mathcal{O}(20)\%$, plus an additional
uncertainty from the definition of the bottom quark mass.


\subsection{Future directions\label{futdir}}

So far we have assumed that our signal in Eq.~(\ref{eq:signal}) mimics the
diphoton signal since the two collimated photons of the light-pseudoscalar decay
are indistinguishable from an isolated photon. However, if the four-photon
final state was discriminated from the diphoton signature, it would be a
strong hint at our
scenario. References~\cite{Ellis:2012zp,Ellis:2012sd,Dasgupta:2016wxw} considered
photon jets (two or more collimated photons)
at hadron colliders. In particular, Ref.~\cite{Dasgupta:2016wxw}
discussed the possibility of photon conversion into $e^+e^-$ pairs and its
discriminating power between photon jets and isolated photons. For a photon
jet, the probability of photon conversion is higher than for a single photon,
and Ref.~\cite{Dasgupta:2016wxw} showed that already several tens of events
are sufficient to discriminate between both hypotheses and a few hundred
events allow for a $5\sigma$ discrimination assuming prompt
photons. However, their conclusions assume a pseudoscalar mass of 1~GeV
and the results are very sensitive to this parameter. For long-lived
pseudoscalars, the discriminating power is reduced since photon conversion
cannot start before the pseudoscalar decay. As a consequence, the
discriminating power becomes worse for increasing lifetimes.  

Apart from the diphoton signal, which is the main motivation of the current study, the NMSSM parameter points discussed here also have additional distinctive features closely related to the diphoton signal. We shortly discuss these collider signatures.

As discussed in the previous section, the light pseudoscalar, $A_1$, has a
small branching fraction of $\lsim1\%$ for decays to electron
pairs. Because of its short life-time it would typically decay promptly
to a highly collimated $e^+e^-$ pair, so-called ``electron jet''. Such
electron jets, prompt or displaced, were searched for by the LHC
experiments as they can appear in many different models of new physics. In
our case, two signatures can appear: two high-$p_T$ electron jets or one
electron jet and an energetic photon. Note that even though we have
suppressed the branching ratio of the SM-like Higgs boson to pseudoscalars
such a signal is also possible apart from the decays of the heavy Higgs
states. For the 125~GeV Higgs an associated production of 
$pp \to h W (\to \ell \nu)$  (where $h$ denotes the SM-like Higgs boson)
was studied by ATLAS~\cite{Aad:2013yqp}. The obtained limit is 
weak and together with the already mentioned suppression of 
${H_1} \to A_1 A_1$
this is an unlikely discovery channel. The searches for the direct  
production of the scalar decaying to two electron jets could provide further constraints,
but the limits have been obtained only for the light SM-like Higgs
boson~\cite{Aad:2014yea,Aad:2015sms}. Nevertheless this signature might become
interesting in the current run once the diphoton signal is firmly
confirmed. Corresponding studies directly applicable to the heavy
Higgs particles have 
also been performed~\cite{CMS:2014hka,Aad:2015rba}. It is interesting to note
that for a hypothetical scenario with a 1~TeV scalar resonance, the
sensitivity of the CMS search~\cite{CMS:2014hka} is in the fb range. While the
discussed 8~TeV searches lack the sensitivity to constrain our scenario now,
they clearly offer interesting prospects for observing electron decay modes of
$A_1$ (possibly accompanied by the photon jet from the opposite decay chain) at the increased center-of-mass energy and the high luminosity run of the
LHC.

Our scenario can also be probed via the ``classic signature'' for additional
heavy neutral Higgs bosons, $pp \to \Phi \to \tau^+\tau^-$, where the
limits are set in the $m_\Phi$--$\tan\beta$ space. Within the
MSSM, assuming the additional Higgs bosons at a mass around 
${\sim }750$~GeV, the (expected) limits on $\tan\beta$ are around 
${\sim }35$ based on Run~I data~\cite{Khachatryan:2014wca,Aad:2014vgg}
(see also~\cite{Bechtle:2015pma}). 
In our NMSSM scenario we have three Higgs bosons with a mass around 750~GeV
contributing to this search channel, $H_2$, $H_3$ and $A_2$, 
where the overall number of $\tau^+\tau^-$ events is roughly 25\% lower than
in the MSSM, mainly due to the decay of $H_{2,3} \to A_1 A_1$. 
Consequently, a similar, but slightly higher limit on $\tan\beta$ can be set
in our NMSSM scenario.
With increasing luminosity this limit could roughly improve to 
$\tan\beta \sim 5$--$10$ at the LHC after collecting 300--3000/fb of integrated luminosity (see also~\cite{Holzner:2014qqs}). 
Therefore, the proposed scenario could eventually lead to an
observable signal in the $\tau^+\tau^-$ searches for heavy Higgs bosons
at the LHC, depending on the details of the scenario (value of
$\tan\beta$, masses of electroweak particles etc.).

Another prediction that arises for parameter points considered in this study
are light higgsinos. With the masses of $100$--$300$~GeV they are well within
the kinematic reach of the LHC. However, the small mass differences,
$\mathcal{O}(10 \gev)$, within the light higgsino sector hinder their
observation at the LHC. If all the non-higgsino SUSY particles are sufficiently far in mass (points {\bf P1}--{\bf P4}, {\bf P8}, {\bf P9}), the decay of the
second neutralino, $\tilde{\chi}^0_2$ proceeds almost exclusively via the
light pseudoscalar $A_1$. With the following significant branching ratio to a
soft $\gamma \gamma$ pair the observation in the soft di- and trilepton
searches~\cite{Khachatryan:2015pot,vanBeekveld:2016hbo} becomes
practically impossible. The radiative production at a high-energy $e^+e^-$
collider remains a valid
possibility though~\cite{Berggren:2013vfa,Moortgat-Picka:2015yla}.

Finally, light smuons are required in order to obtain phenomenologically
viable muon anomalous magnetic moment and to counteract the effects of a
very light pseudoscalar. For our parameter points we fix slepton
masses at $300$--$400$~GeV. While this is close to the existing simplified
model limits, see e.g.~\cite{Aad:2014vma}, in our case due to the significant
branching ratio $\BR(\tilde{\ell}_L \to \tilde{\chi}_1^\pm \nu) \gtrsim 50\%$
these constraints are significantly relaxed. Nevertheless, if the slepton and
higgsino spectra are favorable, the observation in the current LHC run is
plausible.


\section{Conclusions}

We have proposed an NMSSM scenario that can explain the excess in the
diphoton spectrum at $750$~GeV recently observed by ATLAS and CMS. 
In our scenario the heavy neutral (and charged) Higgs bosons have a mass
around ${\sim }750 \gev$, while one light CP-odd Higgs boson has a mass
around the mass of the pion, ${\sim }135 \mev$. 
The $750$~GeV excess is generated by the production of the heavy
neutral CP-even Higgs bosons, which subsequently decay to two light
pseudoscalars. Each of these pseudoscalars then decays, mainly
via the mixing with the $\pi^0$, to a collimated photon pair that
appears as a single photon in the electromagnetic calorimeter.
The mass gap between heavy Higgses of ${\cal O}(20) \gev$ mimics a
large width of the $750 \gev$ peak. Furthermore, the production of the
heavy neutral CP-even Higgs bosons may contain a large component of
$b \bar b$ initial state, thus ameliorating a possible
tension with $8$~TeV data compared to other production modes.
The main virtue of our scenario is that all necessary properties to fit
the signal can be united in a phenomenologically realistic way within as
theoretically simple a model as the NMSSM, without  
e.g.\ requiring additional ad hoc matter.

We derived the NMSSM parameter space in which this scenario can be
realized. It is characterized by a heavy Higgs boson mass scale, $M_A$,
around $750 \gev$. The Yukawa-like couplings $\lambda$ and $\kappa$ are
found to satisfy $\frac{0.4\tan\beta}{1+\tan^2\beta}\lsim\lambda\lsim
\frac{2\sqrt{2}\tan{\beta}}{\sqrt{1+18\tan^2{\beta}+\tan^4{\beta}}}$ and
$\kappa\simeq\frac{\lambda}{2\sin{2\beta}}$. The $\mu$ parameter is
given by $\mu \sim M_A \sin 2\beta$, or 
$2 \frac{\kappa}{\lambda}\mu\simeq750 \gev$. We furthermore find 
$5 \lsim \tan\beta \lsim 15$, $A_{\kappa} \lsim {\cal O}(0.1) \gev$ and
$A_{\lambda}\ll v$.  
Due to this choice of parameters the two neutral heavy CP-even Higgs bosons
are strongly mixed doublet/singlet states and the light CP-odd Higgs boson
can have a mass around $m_{\pi}$. The light CP-even Higgs boson with SM-like
properties can 
have a mass around ${\sim }125 \gev$, mainly by
choosing the scalar top parameters accordingly. The parameter choice
furthermore forbids a large decay rate of the SM-like Higgs boson to the
two light CP-odd states, which would be in contradiction with the LHC
measurements. 

In order to validate our scenario we have chosen nine benchmark points,
all satisfying the above constraints, but with a strong variation within
the allowed intervals. Using state-of-the-art tools, including
higher-order corrections, these points have been analyzed to reproduce
the observed ``excess'' in the diphoton search at the LHC Run-II,
including detector simulation and efficiencies. We have furthermore
checked explicitly that these points fulfill all other experimental
constraints. These include LHC Higgs (and SUSY) searches, Higgs boson
rate measurements, as well as flavor observables and electroweak
precision data. We have shown explicitly that the two collimated photon
pairs would be seen as a single photon each, applying the same settings
as in the ATLAS/CMS analyses. 

Finally, we have analyzed how our scenario can be probed in the upcoming
continued LHC Run-II. Possibly striking features are the absence of any
other relevant decay mode, such as the decay to massive gauge bosons, as
well as an increased rate of photon conversion to electron jets
with respect to the ``simple'' diphoton decay mode, or the distinction of a photon pair
from a single photon. Furthermore, the heavy neutral Higgs bosons should
be visible in the conventional $\tau^+\tau^-$ searches at high
luminosity. Other characteristic features of our scenario are relatively
light higgsinos and possibly sleptons that can be probed at the LHC Run-II.
Using these characteristics, our scenario should be distinguishable from
most other physics scenarios that have been proposed to explain the LHC
diphoton ``excess''.


\section*{Acknowledgements}

F.D.\ wishes to thank S.\ Liebler for his assistance with
\verb|SusHi|. The authors acknowledge useful discussions with 
F.~Jegerlehner
and
A.~Thomas.
J.S.K.\ and K.R.\ have been partially supported by the MINECO (Spain) under contract FPA2013-44773-P; 
Consolider-Ingenio CPAN CSD2007-00042; the Spanish MINECO Centro de
excelencia Severo Ochoa  Program under grant SEV-2012-0249; and by
JAE-Doc program. F.D.\ and S.H.'s work has been supported by CICYT
(grant FPA 2013-40715-P). The work of S.H.\ was supported in part by
Spanish MICINN's Consolider-Ingenio 2010 Program under grant 
MultiDark CSD2009-00064.



\end{document}